# On Algorithms and Complexity for Sets with Cardinality Constraints

## PSPACE and PTIME Logics for Program Analysis

Bruno Marnette  
ENS de Cachan, MPRI  
Paris, France  
bruno@marnette.fr

Viktor Kuncak    Martin Rinard  
MIT Computer Science and Artificial Intelligence Lab  
Cambridge, USA  
{vkuncak,rinard}@csail.mit.edu

## Abstract

Typestate systems ensure many desirable properties of imperative programs, including initialization of object fields and correct use of stateful library interfaces. Abstract sets with cardinality constraints naturally generalize typestate properties: relationships between the typestates of objects can be expressed as subset and disjointness relations on sets, and elements of sets can be represented as sets of cardinality one. In addition, sets with cardinality constraints provide a natural language for specifying operations and invariants of data structures.

Motivated by these program analysis applications, this paper presents new algorithms and new complexity results for constraints on sets and their cardinalities. We study several classes of constraints and demonstrate a trade-off between their expressive power and their complexity.

Our first result concerns a quantifier-free fragment of Boolean Algebra with Presburger Arithmetic. We give a nondeterministic polynomial-time algorithm for reducing the satisfiability of sets with symbolic cardinalities to constraints on constant cardinalities, and give a polynomial-space algorithm for the resulting problem. The best previously existing algorithm runs in exponential space and nondeterministic exponential time.

In a quest for more efficient fragments, we identify several subclasses of sets with cardinality constraints whose satisfiability is NP-hard. Finally, we identify a class of constraints that has polynomial-time satisfiability and entailment problems and can serve as a foundation for efficient program analysis. We give a system of rewriting rules for enforcing certain consistency properties of these constraints and show how to extract complete information from constraints in normal form. This result implies the soundness and completeness of our algorithms.

## 1. Introduction

Program analyses that reason about deep semantic properties are of great value for software development; the value of such analyses is growing with the adoption of language constructs that eliminate low-level program errors. Many deep semantic properties are naturally expressible in fragments of set theory, so constraint solving for such fragments is of interest. This paper presents new algorithms and improved complexity bounds for fragments of set theory. The starting point of our constraints is the boolean algebra of finite (but unbounded) sets.

**Sets in program analysis.** The boolean algebra of finite sets is a fragment of set theory that allows the basic set operations of intersection, union, and complement on sets of uninterpreted elements. Although simple, it turns out that this fragment can express many properties of interest in program analysis. Examples include typestate properties and public interfaces of data structures.

Set specifications generalize typestate properties [29, 26]: the fact that an object $o$ is in the typestate $t$ is represented as the set membership of $o$ in $t$. Through inclusion and disjointness constraints, sets can also express relationships (such as hierarchy or orthogonality) between different typestates. Objects can be represented as sets of cardinality one using a cardinality constraint $|o| = 1$, so set membership reduces to subset. Multiple set memberships can then encode constraints such as $|t| \geq k$ for any constant $k$.

Sets can also provide natural abstractions of container data structures. When a content of a data structure is represented as an abstract set $s$, an operation such as insertion can be characterized by a postcondition $s' = s \cup e$ where $e$ is the set corresponding to the element being inserted. By expressing both typestates and data structure abstractions, sets can be used to combine the results of different analyses operating on the same program. Such an approach allows us to combine the scalability of typestate analysis with the precision of shape analysis and theorem proving [30, 28, 27, 46].

**Sets with cardinality constraints.** The use of the cardinality operator on sets leads to a connection between set algebra operations and integer linear arithmetic, as evidenced, for example, in the condition $|a \cup b| = |a| + |b|$ for disjoint sets $a$ and $b$. It is therefore natural to consider constraints that combine integer linear arithmetic with set algebra operations. These constraints constitute the Quantifier-Free Boolean Algebra with Presburger Arithmetic, or QFBAPA for short — they are the quantifier-free fragment of BAPA constraints whose decision procedure and complexity we have studied in [23, 22]. QFBAPA constraints can be used to verify an invariant such as $|a| = |b|$ which allows us to conclude that if $a$ is nonempty, so is $b$, and therefore it is possible to call an operation that removes an element from $b$. Similarly, if $i$ is an integer variable and $s$ is a set, it is possible to verify an invariant $|s| = i$ stating that an integer $i$ correctly maintains the size of the set $s$. In our experience, specialized decision procedures such as [22] are the only automated technique for deciding with non-trivial cardinality constraints. Currently, however, the complexity of these decision procedures limits their applicability. In this paper we give new algorithms for solving set cardinality constraints; these algorithms provide exponential improvements over existing approaches and make the checking of cardinality constraints in larger formulas more feasible.

Our paper provides a systematic study of constraints on sets in the presence of cardinalities. We study both more expressive and less expressive fragments and demonstrate a trade-off between the

$$
\begin{aligned}
F &::= A \mid F_1 \wedge F_2 \mid F_1 \vee F_2 \mid \neg F \\
A &::= B_1 = B_2 \mid B_1 \subseteq B_2 \mid T_1 = T_2 \mid T_1 \leq T_2 \mid K \text{ dvd } T \\
B &::= s \mid \mathbf{0} \mid \mathbf{1} \mid B_1 \cup B_2 \mid B_1 \cap B_2 \mid B^c \\
T &::= i \mid K \mid T_1 + T_2 \mid K \cdot T \mid |B| \\
K &::= \ldots \mid -2 \mid -1 \mid 0 \mid 1 \mid 2 \mid \ldots
\end{aligned}
$$

**Figure 1.** Quantifier-Free Formulas of Boolean Algebra with Presburger Arithmetic (QFBAPA)

expressive power and the efficiency of the algorithms. The main contributions of our paper are the following:

- **PSPACE algorithm for** QFBAPA. The best previously known algorithms for QFBAPA [23, 22, 45] execute in non-deterministic exponential time, and involve searching for an exponentially large object. In this paper we first give a form of bounded model property that shows that it is possible to replace reasoning about symbolic cardinalities such as $|a| = i \wedge |b| = i$ where $i$ is an integer variable, with guessing sufficiently large constant cardinalities, such as $|a| = 1000 \wedge |b| = 1000$. Moreover, we give a space-efficient algorithm for solving the resulting constraints on sets with large constant cardinalities. This gives a PSPACE decision procedure for QFBAPA and is the first contribution of this paper.

- **A Polynomial-Time Class.** Given that QFBAPA constraints are NP-hard, the question remains whether there are interesting fragments of sets with cardinalities which can be reasoned about in polynomial time. In a quest for such fragments, we identify several features of constraints, each of which leads to NP-hardness. By eliminating these features we have discovered a class (called *i-trees*) that has a polynomial-time satisfiability and entailment (subsumption) problems, while still supporting subset, union, disjointness, and arbitrarily large cardinality constraints. This class can therefore express generalized typestate constraints such as multiple orthogonal classifications into independent or disjoint sets. The identification of this polynomial-time class, and the development of algorithms for testing the satisfiability and subsumption of constraints in this class is the second contribution of this paper. While the resulting algorithms are efficient, the proof of their completeness is somewhat lengthy, and involves characterizations of normal forms of i-trees and the construction of models for i-trees in normal form. We therefore only summarize the main ideas; we refer the reader to the full version of the paper [32] for details. Additional proofs are also included in the Appendix.

We proceed by defining the fragment QFBAPA in Section 2. We present a PSPACE algorithm for QFBAPA in Section 3, defining the simpler CBAC constraints and identifying their NP-complete fragment, CBASC constraints.

## 2. Constraints on Sets with Cardinalities

***Boolean Algebra with Presburger Arithmetic.*** Figure 1 presents the syntax of the constraints studied in this paper, we call these formulas Quantifier-Free Boolean Algebra with Presburger Arithmetic (QFBAPA). QFBAPA constraints contain two kinds of values: integers and sets, each with corresponding applicable operations. The sets are interpreted as subsets of some arbitrarily large finite set. $s$ denotes a set variable, $i$ denotes an integer variable. The symbol $|B|$ denotes the cardinality of the set $B$ and establishes the connection between set and integer terms. MAXC is a special free variable denoting the size of the universal set. If $b$ is a set, $b^c$ denotes its complement. $K$ dvd $T$ denotes that $K$ divides $T$. $K$ denotes constants, encoded in binary: a constant $k$ is encoded using $O(\log k)$ bits. The symbol $A$ in Figure 1 denotes atomic formulas; a literal is an atomic formula or its negation.

A *quantified version* of this language (BAPA) is studied in [23, 22]; where we give an algorithm that establishes a doubly exponential space upper bound on the complexity. Because quantified BAPA subsumes Presburger arithmetic, the doubly exponential nondeterministic time lower bound [15] applies to BAPA as well.

**Preliminaries.** If $S$ is a finite set, $|S|$ denotes the number of elements in $S$. A literal is an atomic formula or its negation. $\mathbb{Z} = \{\ldots, -1, 0, 1, \ldots\}$ is the set of integers, $\mathbb{N} = \{0, 1, \ldots\}$ is the set of natural numbers. $[a..b]$ denotes the set of integers $\{a, a+1, \ldots, b\}$. If $f : A \to B$ is a function and $S \subseteq A$, we define $f[S] = \{f(a) \mid a \in S\}$.

If $A$ is a set, the notation $A^y$ has several potential meanings; the specific meaning should be clear from the context. $A^n$ for $n \in \{1, 2, \ldots, \}$ is the set of vectors $(a_1, \ldots, a_n)$ where $a_j \in A$ for $1 \leq j \leq n$, and $A^{m,n}$ is the set of matrices $[a_{pq}]$ with $m$ rows and $n$ columns with elements $a_{pq}$ for $1 \leq p \leq m$ and $1 \leq q \leq n$. The expression $A^c$ denotes the complement of the set $A$. If $\alpha \in \{0, 1\}$, then $A^\alpha$ denotes $A$ for $\alpha = 1$ and $A^c$ for $\alpha = 0$.

The relation $\equiv$ denotes the equality of the values of metavariables denoting syntactic objects, so if $f_1$ and $f_2$ are formulas, then $f_1 \equiv f_2$ means that they are the same formula. In the context of inclusion diagrams (Section 4), $\equiv$ will denote the semantic equivalence of diagrams (we use $=$ to denote the equality of diagrams).

## 3. A PSPACE Algorithm for QFBAPA

Verification conditions arising in program verification can often be expressed using quantifier-free formulas, so it is natural to examine whether more efficient algorithms exist for QFBAPA constraints. When applied to QFBAPA formulas, existing algorithms run in non-deterministic exponential time (NEXPTIME): the algorithm [45] requires nondeterministically guessing an exponentially large object, whereas the algorithm $\alpha$ from [22] produces an exponentially large quantifier-free Presburger arithmetic formula. The question arises whether there exist algorithms that avoid *non-deterministically guessing* exponentially large objects. We show that this is indeed the case. Namely, we first show that Presburger arithmetic formulas generated by the algorithm $\alpha$ from [22] can in fact be solved in *deterministic* exponential time. Our result reduces QFBAPA to a simpler system of CBAC constraints (shown in Figure 3), then applies a theorem by Papadimitriou [36] in a novel way. This leads to a deterministic EXPTIME decision procedure for QFBAPA satisfiability, which is an improvement on previously existing algorithms. Nevertheless, the question arises whether it is possible to avoid the construction of a non-deterministically large system of equations. It turns out that this is indeed possible: we present an alternating polynomial-time (and therefore, PSPACE) algorithm for QFBAPA. Therefore, it is possible to solve QFBAPA using solvers for quantified boolean formulas [9, 48, 37].

Figures 2 and 4 present our PSPACE algorithm for QFBAPA. The algorithm has two phases.

In the first phase, the non-deterministic polynomial-time algorithm in Figure 2 reduces QFBAPA constraints to a simpler class of constraints. We call these simpler constraints *Conjunctions of Boolean Algebra expressions with Cardinalities* (CBAC). CBAC constraints have a very simple syntactic structure (see Figure 3), but capture the key difficulty in solving QFBAPA: the need to consider exponentially large cardinalities on exponentially many set partitions.

In the second phase, the algorithm in Figure 4 checks the satisfiability of CBAC in alternating polynomial time and therefore in polynomial space. The key insight behind our algorithm is that it is possible to use a divide and conquer approach to avoid explicitly representing all possible regions in the Venn diagram.

Let $f$ be the input QFBAPA formula.

1. Replace each $\mathbb{Z}$-variable with a difference of two $\mathbb{N}$-variables:
   $$C[i_1, \ldots, i_n] \to C[i'_1 - i''_1, \ldots, i'_n - i''_n]$$
   $i'_1, i''_1, \ldots, i'_n, i''_n$ are fresh $\mathbb{N}$-variables

2. Ensure that all set algebra expressions appear within cardinality constraints by normalizing with the following rules:
   $$C[b_1 = b_2] \to C[b_1 \subseteq b_2 \wedge b_2 \subseteq b_1]$$
   $$C[b_1 \subseteq b_2] \to C[|b_1 \cap b_2^c| = 0]$$

3. Eliminate divisibility constraints:
   $C[k \text{ dvd } t] \to C[ki = t]$, $i$ is fresh $\mathbb{N}$-variable

4. Move all cardinality constraints to top level:
   $$C[|b_1|, \ldots, |b_w|] \to f_1 \wedge f_2$$
   where $f_1 \stackrel{def}{\equiv} C[i_1, \ldots, i_w]$
   $f_2 \stackrel{def}{\equiv} |\mathbf{1}|{=}\mathsf{MAXC} \wedge \bigwedge_{j=1}^{w} |b_j|{=}i_j$
   and $i_1, \ldots, i_w$ are fresh $\mathbb{N}$-variables; let $m_1 = w + 1$;

5. Let $p$ be a propositional formula such that $p(a_1, \ldots, a_{m_0}) \equiv f_1$ for atomic formulas $a_1, \ldots, a_{m_0}$. Nondeterministically select the truth value $\alpha_j \in \{0,1\}$ for each atomic formula $a_j$, so that $p(a_1, \ldots, a_{m_0})$ is true. Let $f_{11} \stackrel{def}{\equiv} \bigwedge_{j=1}^{m_0} a_j^{\alpha_j}$.

6. For each conjunct $\neg(t_1{=}t_2)$ in $f_{11}$, non-deterministically replace the conjunct with one of the conjuncts $(t_1 + 1 \leq t_2)$ or $(t_2 + 1 \leq t_1)$.

7. Transform linear integer constraints to normal form:
   $$C[\neg(t_1 \leq t_2)] \to C[t_2 + 1 \leq t_1]$$
   $$C[t_1 \leq t_2] \to C[t_1 - t_2 + i = 0]$$
   $$C[t_1 = t_2] \to C[\sum_{j=1}^{n} c_j i_j = k]$$

8. Let $n_0$ be the number of $\mathbb{N}$-variables in the entire formula. The resulting system is of the form:
   $$Av = d \wedge \bigwedge_{j=1}^{m_1} |b_j| = i_{p_j}$$
   where $A \in \mathbb{Z}^{m_0,n_0}$, $d \in \mathbb{Z}^{m_0}$, and $v = (i_1, \ldots, i_{n_0})$ where each $i_j$ is an $\mathbb{N}$-variable and $1 \leq p_1, \ldots, p_{m_1} \leq m_1$ are variables denoting cardinalities of sets. Let $S$ be the total number of set variables in $b_1, \ldots, b_{m_1}$. Let $m = m_0 + m_1$, $n = \max(n_0, 2^S)$,
   $a = \max(\{1\} \cup \{|a_{pq}| \mid 1 \leq p \leq n_0, 1 \leq q \leq m_0\}$
   $\cup \{|d_q| \mid 1 \leq q \leq m_0\})$
   where $A = [a_{pq}]_{pq}$, and $d = (d_1, \ldots, d_{m_0})$, and let $M = n(ma)^{2m+1}$.

9. Non-deterministically select a vector $k = (k_1, \ldots, k_{n_0})$ where $k_j \in \{0, 1, \ldots, M\}$ for $1 \leq j \leq n_0$, such that $Ak = d$.

10. Call CBAC decision procedure on $\bigwedge_{j=1}^{m_1} |b_j| = k_{p_j}$. If there exists a solution, then report the formula satisfiable.

**Figure 2.** An NP Algorithm for Reducing QFBAPA Constraints to CBAC constraints of Figure 3

$$F ::= |B|{=}K \mid F_1 \wedge F_2$$
$$B ::= s \mid \mathbf{0} \mid \mathbf{1} \mid B_1 \cup B_2 \mid B_1 \cap B_2 \mid B^c$$
$$K ::= 0 \mid 1 \mid 2 \mid \ldots$$

**Figure 3.** Conjunctions of Boolean Algebra expressions with Cardinalities (CBAC)

Given a CBAC constraint
$$\sum_{j=1}^{m_1} |b_j| = k_j$$
where the free set variables of $b_1, \ldots, b_{m_1}$ are among $s_1, \ldots, s_S$, run CBAC-check$([], d)$ with $d = (k_1, \ldots, k_{m_1})$.

proc CBAC-check$([v_1, \ldots, v_n], d)$ returns result
    where $v_1, \ldots, v_n$, result $\in \{0, 1\}$; $d \in \mathbb{N}^{m_1}$
  if $(n < S)$ then
    existentially choose $d_0, d_1 \in \mathbb{N}^{m_1}$ such that $d_0 + d_1 = d$;
    universally do
      $r_1 =$ CBAC-check$([v_1, \ldots, v_n, 0], d_0)$ and
      $r_2 =$ CBAC-check$([v_1, \ldots, v_n, 1], d_1)$;
    return $r_1 \wedge r_2$;
  else
    let $p_j =$ eval$(b_j, [s_1 \mapsto v_1, \ldots, s_S \mapsto v_S])$
        for all $(1 \leq j \leq m_1)$;
    $J_0 = \{d_j \mid p_j = 0\}$;
    $J_1 = \{d_j \mid p_j = 1\}$;
    return $J_0 \subseteq \{0\} \wedge |J_1| \leq 1$.

proc eval$(b, \alpha)$ returns result
    where $b$ : Boolean Algebra formula
        $\alpha : \{s_1, \ldots, s_S\} \to \{0, 1\}$
        result $\in \{0, 1\}$
  treating $b$ as a propositional formula,
  return the value of $b$ under assignment $\alpha$.

**Figure 4.** An Alternating Polynomial-Time (and PSPACE) Algorithm for Checking the Satisfiability of CBAC Constraints

We next discuss our algorithm in more detail and argue that it is correct. We begin with the description of the steps of the algorithm in Figure 2, which reduces symbolic cardinalities to large constant cardinalities.

*1. Non-negative integers.* To simplify the later steps, the first step makes all integer variables range over non-negative integers $\mathbb{N}$, by replacing each integer variable $i$ with a difference $i_1 - i_2$ of fresh non-negative integer variables $i_1, i_2$.

*2,3. Eliminating set equality and subset, and integer divisibility.* The next step converts set equality and set subset into cardinality constraints. This step helps the later separation between the boolean algebra part and the integer linear arithmetic part. We then eliminate any divisibility relations using multiplication and a fresh variable.

*4. Flattening.* The next step separates the formula into the boolean algebra part, denoted $f_1$ and the integer linear arithmetic part, denoted $f_2$. This step simply amounts to naming the cardinality of each set by a fresh integer variable.

*5,6. From quantifier-free formulas to conjunctions.* An obvious source of NP-completeness of QFBAPA is the presence of arbitrary propositional combinations of atomic formulas. An effective way of dealing with propositional combinations is to enumerate the satisfying assignments of the propositional formula using a SAT solver, and then solve the conjunctions of literals [16, 17]. Steps 5 and 6 of the non-deterministic algorithm in Figure 2 are an abstract description of such procedure. The goal of step 6 is to eliminate disequalities, which involve non-deterministic choice between the two inequalities.

***7. Normal form for integer constraints.*** The algorithm eliminates the remaining negations of atomic formulas and transforms linear constraints into normal form $Av = d$.

***8,9,10. Estimating sizes of integer variables.*** The resulting system contains linear integer equations of the form $\sum_{j=1}^{n} c_j i_j = k$, and set cardinality constraints of the form $|b| = i$. The algorithm computes an upper bound $M$ on integer variables in any potential solution of the system, using several parameters: the number of conjuncts $n$, the number of integer variables $n_0$ and the number of set variables $\mathcal{S}$. The computation of the upper bound is based on an observation that the satisfiability of the conjunction of constraints $|b| = i$ can be reduced to the satisfiability of equations of the form $\sum_{j=1}^{p} l_j = i$, where variables $l_j$ denote sizes of set partitions (regions in Venn diagram) whose union is the set $b$; this is a specialization of the idea in [22] to the case of quantifier-free formulas.

Let $s_1, \ldots, s_\mathcal{S}$ be all set variables appearing in formula and consider a constraint $|b| = i$. Consider all partitions $\bigcap_{j=1}^{n} s_j^{\alpha_j}$ for $\alpha_j \in \{0, 1\}$. For each such partition $b_p$, introduce a fresh $\mathbb{N}$-variable $l_p$, which denotes the cardinality of cube $b_p$. Then consider a constraint of the form $|b| = i$. Each set is a union of regions in the Venn diagram (by the disjunctive normal form theorem) so suppose that $b = b_{p_1} \cup \ldots \cup b_{p_a}$. Then replace the term $|b| = i$ with the $\sum_{q=1}^{a} l_{p_q} = i$. We use the term "CBAC linear equations" to denote a system of linear equations resulting from the constraints $|b| = i$ as described above.

As a result, we obtain a system of $m_0 + m_1$ linear equations over non-negative integers, where $m_0$ equations have a polynomial number of variables, and $m_1$ equations (CBAC linear equations) have exponentially many variables. It is easy to see that there exists a surjective mapping of solutions of the original constraints on sets onto solutions of the resulting linear equations (the mapping computes the cardinality of each Venn diagram). Therefore, the original system is satisfiable if and only if the resulting equations are satisfiable. Moreover, we have the following fact.

FACT 1 (Papadimitriou [36]). *Let $A$ be an $m \times n$ integer matrix and $b$ an $m$-vector, both with entries from $[-a..a]$. Then the system $Ax = b$ has a solution in $\mathbb{N}^m$ if and only if it has a solution in $[0..M]^m$ where $M = n(ma)^{2m+1}$.*

Fact 1 implies that the estimate $M$ computed in step 8 of the algorithm in Figure 2 is a correct upper bound. Using this estimate, step 9 of the algorithm non-deterministically guesses the values of all integer variables such that the original linear equations $Ax = d$ are satisfied. All this computation can be performed in nondeterministic polynomial time, and (unlike [22]), does not involve constructing explicitly a system with exponentially many equations. Having picked the values of integer variables, including the variables $i$ on the right hand side of constraints $|b| = i$, we obtain a conjunction of constraints of the form $|b| = k$ where $k$ is a constant whose binary representation has polynomially many bits—these are precisely the CBAC constraints in Figure 3. We have therefore shown the following.

LEMMA 1. *The algorithm in Figure 2 reduces in non-deterministic polynomial time the satisfiability of a QFBAPA formula to the satisfiability of CBAC formulas.*

It remains to find an algorithm for CBAC constraints.

***A PSPACE algorithm for CBAC.*** One correct way to solve CBAC constraints is to solve the associated CBAC linear equations. This system has exponentially many variables, each of which can take any value from $[0..M]$. Therefore, guessing the values of each of these variables can be done in non-deterministic exponential time; similar approaches not based on equations also require guessing exponentially large objects [45]. Note, however, that there are only polynomially many CBAC linear equations. Using the idea of the proof [36, Corollary 1], we can therefore show that a dynamic programming algorithm can be used to solve the system in polynomial time. In fact, we can use the dynamic programming algorithm from the proof of [36, Corollary 1]. Instead of fixing the size of the equations $m_1$ to be constant, we simply observe that $m_1$ is polynomial in the size of the input, whereas the number of variables is singly exponential. The bound $M$ therefore yields a singly exponential deterministic time dynamic programming algorithm for CBAC. While this is better than existing results, we show that an even better result is achievable.

Clearly, any algorithm that explicitly constructs CBAC equations will require at least exponential time and space. Our solution is therefore to adapt the dynamic programming algorithm to a divide and conquer approach that always represents the equations in terms of their original, polynomially sized, boolean algebra expression. Such an algorithm runs in alternating polynomial time, consuming polynomial space, and is presented in Figure 4. To see the idea of our PSPACE algorithm, consider the CBAC linear system of equations written in the vector form: $\sum_{j=1}^{2^p} a_j l_j = d$ where $d$, $a_j$ are vectors and $l_j$ are the variables for $1 \leq j \leq 2^p$. The algorithm guesses the vectors $d_0, d_1 \in \mathbb{N}^m$ such that $d_0 + d_1 = d$, and recursively solves two equations:

$$\sum_{j=1}^{2^{p-1}-1} a_j l_j = d_0 \quad \wedge \quad \sum_{j=2^{p-1}}^{2^p} a_j l_j = d_1$$

This algorithm creates an OR-AND tree whose search gives the answer to the original problem. A position in the tree is given by the propositional assignment $[v_1, \ldots, v_n]$ to boolean variables. Each leaf in the tree is given by a complete assignment $[v_1, \ldots, v_\mathcal{S}]$ to set variables. Note that we never need to explicitly maintain the system during the divide phase of the algorithm, it suffices to determine in the leaf case $p = 0$ whether the coefficient $a_j$ is 0 or 1. The algorithm does this by simply evaluating each Boolean algebra expression $b$ for the assignment $[v_1, \ldots, v_\mathcal{S}]$.

THEOREM 1. *The algorithm in Figure 4 checks the satisfiability of CBAC constraints in PSPACE. The algorithm given by Figures 2 and 4 checks the satisfiability of QFBAPA constraints in PSPACE.*

Theorem 1 improves the existing algorithms for QFBAPA from both a complexity theoretic and an implementation viewpoint. A deterministic realization of previous NEXPTIME algorithms runs in doubly exponential worst-case time and requires exponential space; a deterministic realization of our new algorithm runs in singly exponential time and consumes polynomial space. Previous algorithms would require running a constraint solver such as a SAT solver [47] on an exponentially large constraint; the new algorithm can be solved by running a quantified boolean algebra solver [48] on a polynomially large constraint.

***NP fragments of CBAC.*** We have seen that both CBAC and QFBAPA constraints are in PSPACE. Both of these classes of constraints are NP-hard, because the constraint $|b| = 1$ is satisfiable iff $b$ is corresponds to a satisfiable propositional formula. Moreover, Lemma 1 shows that QFBAPA constraints are in NP iff CBAC constraints are in NP. For some subclasses of CBAC constraints we can indeed show membership in NP. Define conjunctions of boolean algebra expressions with *small* cardinalities, denoted CBASC, to be the same as CBAC but with constant integers encoded in *unary* notation, where an integer $x$ is represented in space $O(x)$ as opposed to $O(\log x)$; such encoding can therefore be exponentially less compact.

LEMMA 2. *The satisfiability of* CBASC *constraints is NP-complete.*

CBASC solutions are NP-hard because $|b| = 1$ is a CBASC constraint. One way to prove membership in NP is to observe that CBASC is subsumed by the language of set-valued fields which was proven to be in NP [24, 25] by reduction to the universal class of first-order logic formulas, which has the small model property [7, Page 258]. Another way is to consider the notion of *sparse solutions* of CBAC linear equations. An $M$-sparse solution is a solution to CBAC linear constraints with at most $M$ non-zero elements. An $M$-sparse solution to CBAC linear constraints with $2^S$ variables can be encoded as an $M$-tuple of pairs $([v_1, \ldots, v_S], k)$ where the propositional assignment $[v_1, \ldots, v_S]$; encodes one of the $2^S$ integer variables, and $k$ specifies the value of that integer variable. This encoding is polynomial in $M\mathcal{S}w$ where $w$ is the number of bits for representing the largest component of the solution. For any CBAC linear constraint $\bigwedge_{j=1}^{m} |b_j| = k_j$, each solution is $M$-sparse where $M = \max(k_1, \ldots, k_m)$. For CBASC solutions, $M$ is polynomial in the size of the CBASC representation because each $k_i$ is encoded in unary, so sparse solutions can be guessed in polynomial time. This proves that CBASC constraints are in NP.[1]

## 4. Inclusion Diagrams

This section introduces inclusion diagrams (i-diagrams), a graph representation of CBAC constraints. Figure 5 shows a formula with sets and cardinalities and an equivalent i-diagram. I-diagrams allow us to naturally describe fragments of CBAC constraints and the algorithms for checking satisfiability and subsumption of these fragments. The basic idea of i-diagrams is to represent the subset partial order using a graph where sets are annotated with cardinalities, and then indicate the disjointness and union relations by constraints on direct subsets of a set. To efficiently represent equal sets, the nodes in the i-diagram stand not for set names, but for collections of set names that are guaranteed to be equal. Finally, we associate uninterpreted predicates with collections of nodes, representing the fact that elements of given sets satisfy the properties given by the predicate. The uninterpreted predicates illustrate a way to combine i-diagram representations with other constraints.

DEFINITION 1 (i-diagrams). *We fix a finite set* SN *of Set-Names, and a finite set* PN *of predicate names. We denote by* $\text{PN}^{\pm}$ *the set of* atoms $\{+P, -P \mid P \in \text{PN}\}$.
*An* i-diagram *(Inclusion-Diagram) is either the* null-diagram $\perp_d$ *or a tuple* $(\mathbb{S}, \emptyset_d, \text{Sons}, \text{Split}, \text{Comp}, \text{CInf}, \text{CSup}, \Phi)$ *such that:*

- $\mathbb{S} \subseteq \mathcal{P}(\text{SN})$ *is a partition of* SN *containing (nonempty) equivalence classes of set names that are guaranteed to be equal, with* $\emptyset_d \in \mathbb{S}$ *the equivalence class corresponding to names of sets whose interpretation is the empty set $\emptyset$;*
- $\text{Sons} : \mathbb{S} \to \mathcal{P}(\mathbb{S})$ *represents subset relation;*
  *we define* $S \rightsquigarrow S' \overset{def}{\iff} S \in \text{Sons}(S')$*; then* $(\mathbb{S}, \rightsquigarrow)$ *is a graph, so we call elements of $\mathbb{S}$ nodes, and the elements of $\rightsquigarrow$ edges; we write $\overset{*}{\rightsquigarrow}$ for the transitive closure of $\rightsquigarrow$;*
- $\text{Split}, \text{Comp} : \mathbb{S} \to \mathcal{P}(\mathcal{P}(\mathbb{S}))$ *represent disjointness and completeness of set inclusions; if $S$ is a node, then $\text{Split}(S)$ is a set of* split views*, where each view is a nonempty set of sons that*

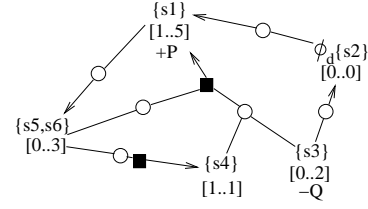

$\mathcal{D}$ is such that $\text{CInf}(\{s_1\}) = 1$, $\text{CSup}(\{s_1\}) = 5$, $\text{Sons}(\{s_1\}) = \{\{s_5, s_6\}, \{s_4\}, \{s_3\}\}$, $\text{Comp}(\{s_1\}) = \{\{\{s_5, s_6\}, \{s_4\}, \{s_3\}\}\}$ $\text{Split}(\{s_1\}) = \{\{\{s_5, s_6\}\}, \{\{s_4\}, \{s_3\}\}\}$, $\Phi(\{s_1\}) = \{+P\}$
and is equivalent to

$s_2 = \emptyset \ \wedge \ s_5 = s_6 \ \wedge$
$s_2 \cup s_3 \cup s_4 \cup s_5 \subseteq s_1 \ \wedge \ s_1 \subseteq s_5 \ \wedge \ s_3 \subseteq s_2 \ \wedge \ s_5 \subseteq s_4$
$s_3 \cap s_4 = \emptyset \ \wedge \ s_1 \subseteq s_3 \cup s_4 \cup s_5 \ \wedge \ s_4 \subseteq s_5$
$1 \leq |s_1| \leq 5 \ \wedge \ |s_4| = 1 \ \wedge \ |s_5| \leq 3 \ \wedge \ |s_3| \leq 2 \ \wedge$
$\forall x \in s_1.\ P(x) \ \wedge \ \forall x \in s_3.\ \neg Q(x)$

**Figure 5.** An example i-diagram $\mathcal{D}$ and an equivalent formula

*represent pairwise disjoint sets, and* $\text{Comp}(S)$ *is a set of* complete views *each of which is a set of nodes that represent sets whose union is equal to the father; we require*

$$\bigcup \text{Split}(S) = \text{Sons}(S)$$
$$\bigcup \text{Comp}(S) \subseteq \text{Sons}(S)$$

*for all $S \in \mathbb{S}$;*
- $\text{CInf}, \text{CSup} : \mathbb{S} \to \mathbb{N}$ *specify lower and upper bounds on the cardinality of sets;*
- $\Phi : \mathbb{S} \to \mathcal{P}(\text{PN}^{\pm})$ *maps nodes to the uninterpreted unary predicates and their negations that are true for all sets of a node.*

To avoid confusion between set names, nodes (sets of set names), and views (sets of nodes), we use lowercase letters $s, s_i, s'$ to denote set names, uppercase letters $S, S_i, S'$ to denote nodes, and letters $Q, C$ to denote views and sets of nodes in general. When $\mathcal{D} \neq \perp_d$ is a diagram, unless otherwise stated, we name its components $\mathbb{S}, \emptyset_d, \text{Sons}, \text{Split}, \text{Comp}, \text{CInf}, \text{CSup}, \Phi$, and similarly we name the components of $\mathcal{D}'$ as $\mathbb{S}', \emptyset_d', \text{Sons}', \text{Split}', \text{Comp}', \text{CInf}', \text{CSup}', \Phi'$.

In a graphical representation of an i-diagram, we represent each element $S \in \mathbb{S}$ where $S = \{s_1, \ldots, s_n\}$ using underlying sets $\{s_1, \ldots, s_n\}$. We represent inclusion $S_1 \rightsquigarrow S_2$ by an arrow from $S_1$ to $S_2$. We represent a split view $Q \in \text{Split}(S)$ where $Q = \{S_1, \ldots, S_n\}$ with a circle connected with undirected edges to $S_1, \ldots, S_n$ and an arrow leading to $S$. We represent a complete view similarly, using a filled square instead of a circle. For each node $S \in \mathbb{S}$ we indicate its cardinality bounds by annotating the node with $[a..b]$ where $a = \text{CInf}(S)$, $b = \text{CSup}(S)$. We represent $\Xi(S) = \{\pm P_1, \ldots, \pm P_n\}$ by annotating $S$ with $\pm P_1, \ldots, \pm P_n$. We represent $\emptyset_d = \{s_1, \ldots, s_n\}$ by annotating the node $\{s_1, \ldots, s_n\}$ with $\emptyset_d$.

DEFINITION 2 (Semantics of i-diagrams). *An* interpretation *of* SN *and* PN *is a triple* $(\Delta, \alpha, \Xi)$ *where*

- $\Delta$ *is a finite set (the universe);*
- $\alpha : \text{SN} \to \mathcal{P}(\Delta)$ *specifies the values of sets;*
- $\Xi : \text{PN} \to \mathcal{P}(\Delta)$ *specifies the values of unary predicates;*

*An interpretation $I$ is a* model *for an i-diagram $\mathcal{D}$, denoted $I \models \mathcal{D}$, iff $\forall s \in \emptyset_d.\alpha(s) = \emptyset$, and for all $S \in \mathbb{S}$ where $S = \{s_1, \ldots, s_n\}$, the following conditions hold:*

---

[1] Sparse solutions are interesting for general CBAC constraints as well. As of yet we have no example of a CBAC constraint whose associated CBAC equation system is satisfiable but has no sparse solutions; moreover, we can generalize the notion of sparse solutions to solutions representable using binary decision diagrams [8] while preserving polynomial-time verifiability.

- $\alpha(s_1) = \ldots = \alpha(s_n)$;
  accordingly, define $\overline{\alpha}(S) \stackrel{def}{=} \alpha(s_1) = \ldots = \alpha(s_n)$
- $\mathsf{CInf}(S) \leq |\overline{\alpha}(S)| \leq \mathsf{CSup}(S)$
- $\forall P.\ (+P) \in \Phi(S) \Rightarrow \overline{\alpha}(S) \subseteq \Xi(P)$
- $\forall P.\ (-P) \in \Phi(S) \Rightarrow \overline{\alpha}(S) \subseteq \Xi(P)^c$
- $\forall S' \in \mathsf{Sons}(S).\ \overline{\alpha}(S') \subseteq \overline{\alpha}(S)$
- $\forall Q \in \mathsf{Split}(S).\ \forall S_1, S_2 \in Q.\ S_1 \neq S_2 \Rightarrow \overline{\alpha}(S_1) \cap \overline{\alpha}(S_2) = \emptyset$
- $\forall Q \in \mathsf{Comp}(S).\ \overline{\alpha}(S) \subseteq \bigcup_{S_1 \in Q} \overline{\alpha}(S_1)$

*We use the standard notions of satisfiability, subsumption (entailment), and equivalence:*

$$\begin{array}{lll} \mathcal{D} \text{ is satisfiable} & \Longleftrightarrow & \exists I.\ I \models \mathcal{D} \\ \mathcal{D}' \models \mathcal{D} & \Longleftrightarrow & \forall I.\ I \models \mathcal{D}' \Rightarrow I \models \mathcal{D} \\ \mathcal{D}' \equiv \mathcal{D} & \Longleftrightarrow & \mathcal{D}' \models \mathcal{D} \wedge \mathcal{D} \models \mathcal{D}' \end{array}$$

DEFINITION 3 (Explicit Disjointness).
*We write* $\mathsf{disj}_{\mathcal{D},S_0}(S_1, S_2)$ *as a shorthand for*

$$S_1 \neq S_2 \wedge \exists Q \in \mathsf{Sons}(S_0).\ S_1, S_2 \in Q$$

*and we say that* $S_1, S_2$ *are* explicitly disjoint, *and we write* $\mathsf{disj}^*_{\mathcal{D}}(S_1, S_2)$ *iff*

$$\exists S'_1, S'_2, S_0 \in \mathbb{S}, S_1 \stackrel{*}{\leadsto} S'_1 \wedge S_2 \stackrel{*}{\leadsto} S'_2 \wedge \mathsf{disj}_{\mathcal{D},S_0}(S'_1, S'_2)$$

LEMMA 3. *I-diagrams have the same expressive power as* CBAC *constraints.*

By "same expressive power" we here mean that there is a natural pair of mappings between the models of i-diagrams and solutions to CBAC constraints.

Because nodes in i-diagrams are collections of set names, we can define the following operations.

DEFINITION 4 (Factor-i-diagram). *Let* $\rho \subseteq \mathbb{S} \times \mathbb{S}$ *be an equivalence relation on nodes. We define* $\mathcal{D}/\rho$ *as follows. Define* $\perp_d/\rho = \perp_d$. *Let* $\mathcal{D} = (\mathbb{S}, \emptyset_d, \mathsf{Sons}, \mathsf{Split}, \mathsf{Comp}, \mathsf{CInf}, \mathsf{CSup}, \Phi)$. *We define* $\mathcal{D}/\rho = \mathcal{D}' = (\mathbb{S}', \mathsf{Sons}', \mathsf{Split}', \mathsf{Comp}', \mathsf{CInf}', \mathsf{CSup}', \Phi')$ *as follows. Define* $h$ *so that if* $\{S_1, \ldots, S_n\}$ *is the equivalence class of* $S$ *under* $\rho$, *then* $h(S) = S_1 \cup \ldots \cup S_n$. *If* $Q \subseteq \mathbb{S}$, *define* $h[Q] = \{h(S) \mid S \in Q\}$. *Then let* $\mathbb{S}' = h[\mathbb{S}]$. *Consider* $S' \in \mathbb{S}'$. *Both* $\mathbb{S}$ *and* $\mathbb{S}'$ *are partitions, and given* $S' \in \mathbb{S}'$ *there is a unique set* $\{S_1, \ldots, S_n\} \subseteq \mathbb{S}$ *such that* $S' = S_1 \cup \ldots \cup S_n$. *Then define:*

$$\begin{aligned} \mathsf{CInf}'(S') &= \max(\mathsf{CInf}(S_1), \ldots, \mathsf{CInf}(S_n)) \\ \mathsf{CSup}'(S') &= \min(\mathsf{CSup}(S_1), \ldots, \mathsf{CSup}(S_n)) \\ \mathsf{Sons}'(S') &= h[\mathsf{Sons}(S_1) \cup \ldots \cup \mathsf{Sons}(S_n)] \\ \Phi'(S') &= \Phi(S_1) \cup \ldots \cup \Phi(S_n) \\ \mathsf{Split}'(S') &= \{h[Q] \mid Q \in \mathsf{Split}(S_1) \cup \ldots \cup \mathsf{Split}(S_n)\} \\ \mathsf{Comp}'(S') &= \{h[Q] \mid Q \in \mathsf{Comp}(S_1) \cup \ldots \cup \mathsf{Comp}(S_n)\} \end{aligned}$$

DEFINITION 5 (Merge). *For any i-diagram* $\mathcal{D}$ *we define the i-diagram* $D[\mathsf{Merge}(Q)] \stackrel{def}{=} \mathcal{D}/\rho$ *for the equivalence relation* $\rho = \{(S_1, S_2) \mid S_1, S_2 \in Q\} \cup \{(S, S) \mid S \in \mathbb{S}\}$

In the sequel we impose the following restrictions on the form of i-diagrams.

DEFINITION 6 (Simple Diagrams). *A diagram is* $\mathcal{D}$ *is* simple *iff* $\mathcal{D} = \emptyset_d$ *or all of the following conditions hold for all* $S \in \mathbb{S}$:

a) $(\mathbb{S}, \leadsto)$ *has no cycles, in particular* $S \notin \mathsf{Sons}(S)$

b) $\emptyset_{\mathcal{D}} \notin \mathsf{Sons}(S)$

c) $\emptyset \notin \mathsf{Split}(S) \wedge \emptyset \notin \mathsf{Comp}(S)$

d) $\forall Q, Q'.\ Q \in \mathsf{Split}(S) \wedge Q' \subsetneq Q \Rightarrow Q' \notin \mathsf{Split}(S)$

e) $\forall Q, Q'.\ Q \in \mathsf{Comp}(S) \wedge Q' \supsetneq Q \Rightarrow Q' \notin \mathsf{Comp}(S)$

f) $\mathsf{CSup}(\emptyset_d) = 0, \mathsf{Sons}(\emptyset_d) = \Phi(\emptyset_d) = \emptyset$

proc **Simplify**$(\mathcal{D})$ :

1. use fixpoint iteration to compute $\rho$ as the smallest equivalence relation such that:

   1.1. $S_1 \stackrel{*}{\leadsto} S_2 \wedge S_2 \stackrel{*}{\leadsto} S_1 \Rightarrow (S_1, S_2) \in \rho$

   1.2. $(S, \emptyset_d) \in \rho \wedge S_1 \stackrel{*}{\leadsto} S \Rightarrow (S_1, \emptyset_d) \in \rho$

   1.3. $\emptyset \in \mathsf{Comp}(S) \Rightarrow (S, \emptyset_d) \in \rho$

   1.4. $\mathsf{disj}_{\mathcal{D},S_0}(S_1, S_2) \wedge (S_1, S_2) \in \rho \Rightarrow (S_2, \emptyset_d) \in \rho$

   1.5. $\mathsf{disj}_{\mathcal{D},S_0}(S_1, S_2) \wedge (S_0, S_1) \in \rho \Rightarrow (S_2, \emptyset_d) \in \rho$

   1.6. $Q \in \mathsf{Comp}(S_1) \wedge (\forall S \in Q.(S, \emptyset_d) \in \rho) \Rightarrow (S_1, \emptyset_d) \in \rho$

2. $\mathcal{D} := \mathcal{D}/\rho$

3. $\begin{bmatrix} \mathsf{Split}(S) \leftarrow \{Q - \{\emptyset_d\} | Q \in \mathsf{Split}(S), S \notin Q\} \\ \mathsf{Comp}(S) \leftarrow \{Q - \{\emptyset_d\} | Q \in \mathsf{Comp}(S), S \notin Q\} \\ \mathsf{Sons}(S) \leftarrow \mathsf{Sons}(S) - \{\emptyset_d, S\} \end{bmatrix}_{S \in \mathbb{S}}$

4. $\begin{bmatrix} \mathsf{Split}(S) \leftarrow \mathsf{Split}(S) - \{\emptyset\} \\ \quad -\{Q \mid \exists Q' \in \mathsf{Split}(S), Q' \supsetneq Q\} \\ \mathsf{Comp}(S) \leftarrow \mathsf{Comp}(S) - \{\emptyset\} \\ \quad -\{Q \mid \exists Q' \in \mathsf{Comp}(S), Q' \subsetneq Q\} \end{bmatrix}_{S \in \mathbb{S}}$

5. $\begin{bmatrix} \mathsf{CSup}(\emptyset_d) \leftarrow 0 \\ \Phi(\emptyset_d) \leftarrow \emptyset \\ \mathsf{Sons}(\emptyset_d) \leftarrow \emptyset \\ \mathsf{Comp}(\emptyset_d) \leftarrow \emptyset \\ \mathsf{Split}(\emptyset_d) \leftarrow \emptyset \end{bmatrix}$

6. return $\mathcal{D}$

Where $[a \leftarrow b]$ denotes the result of updating the component $a$ of i-diagram $\mathcal{D}$ with value $b$.

**Figure 6.** Polynomial-time algorithm **Simplify** to compute an equivalent simple i-diagram

Simplicity eliminates redundancy from diagrams, but does not restrict their expressive power, as the following lemma shows.

LEMMA 4. *For every i-diagram* $\mathcal{D}$ *we can obtain an equivalent simple i-diagram using the polynomial-time algorithm* **Simplify** *in Figure 6.*

## 5. Sources of NP Hardness and Definition of I-Trees

The satisfiability of i-diagrams is NP-hard because i-diagrams have the same expressive power as CBAC constraints. We have observed that the general directed acyclic graph structure of i-diagrams allows us to encode NP-complete problems; this motives the following two restrictions.

DEFINITION 7.
*An i-diagram* $\mathcal{D}$ *is* **tree shaped** *iff*
$(\mathbb{S}, \leadsto)$ *is a tree (with an additional isolated node* $\emptyset_d$)

*An i-diagram* $\mathcal{D}$ *has* **independent views** *iff*
*for all* $Q_1, Q_2 \in \mathsf{Split}(S) \cup \mathsf{Comp}(S)$ *at least one of the following two conditions holds:*

- $Q_1 \cap Q_2 = \emptyset$
- $Q_1 \in \mathsf{Split}(S) \wedge Q_2 \in \mathsf{Comp}(S) \wedge Q_1 \subseteq Q_2$.

Recall that, by Lemma 4, it suffices to consider i-diagrams with acyclic graphs of the subset relation. The tree shape condition is then a natural next restriction on the structure of i-diagrams. However, due to the presence of Split and Comp, the tree shape condition by itself does not reduce the expressive power of i-diagrams, and further restrictions are necessary. The independent views con-

*On Algorithms and Complexity for Sets with Cardinality Constraints*      6      *2018/11/20*

dition extends the tree condition to the entire graphical representation of i-diagrams, including the circles and squares that represent Split and Comp views. The conjunction of these two conditions can be expressed by saying that the graphical representation of i-diagram is a tree.

REMARK 1. We can express the combination of the conditions: being simple, being tree shaped, and having independent views by saying that there are only four kinds of edges in the corresponding graphical representation:[2]

- from an element $S \in \mathbb{S} - \{\emptyset_d\}$ to a circle
- from a circle to a square, indicating that all nodes of a split view belong to a complete view
- from a circle to an element $S \in \mathbb{S} - \{\emptyset_d\}$,
- from a square to an element $S \in \mathbb{S} - \{\emptyset_d\}$.

Unfortunately, the restrictions on tree shape and independent views are not sufficient to guarantee a polynomial-time decision procedure in the presence of predicates associated with nodes. The reason is that the ability to encode disjointness of arbitrary sets leads to NP-hardness, yet even with tree structure and independent views it is possible to assert that two arbitrary sets $S_1$ and $S_2$ are disjoint by letting $(+P) \in \Phi(S_1)$ and $(-P) \in \Phi(S_2)$ for some uninterpreted predicate $P$. A simple way to avoid this problem is to require that $\Phi$ contains only positive atoms $(+P)$. A more flexible restriction is the following.

DEFINITION 8. *An i-diagram $\mathcal{D}$ has **independent signatures** iff for every pair of distinct nodes $S_1, S_2$ such that $(-P) \in \Phi(S_1)$ and $(+P) \in \Phi(S_2)$ for some $P \in \mathsf{PN}$, at least one of the following two conditions holds:*

1. *$S_1$ and $S_2$ are* explicitly disjoint, *that is,* $\mathsf{disj}^*_{\mathcal{D}}(S_1, S_2)$
2. *$S_1$ and $S_2$ have* compatible signatures, *that is, there exists a node $S$ such that*

$$S_1 \stackrel{*}{\rightsquigarrow} S \land S_2 \stackrel{*}{\rightsquigarrow} S \land$$
$$\mathsf{Sig}(S_1) \cap \mathsf{Sig}(S_2) \subseteq \mathsf{Sig}(S)$$

*where* $\mathsf{Sig}(S) = \{P \mid (+P) \in \Phi(S) \lor (-P) \in \Phi(S)\}$.

The independent signatures condition ensures that any disjointness conditions are either 1) a result of the fact that the ancestors of $S_1$ and $S_2$ are explicitly stated as disjoint, or 2) a result of a contradictory predicate assignment (the case when $S_1$ and $S_2$ have compatible signatures, so there exists a parent that resolves which of $(+P)$ or $(-P)$ hold for both $S_1$ and $S_2$).

The discussion above leads to the definition of i-trees, for which we will give polynomial-time algorithms for satisfiability and subsumption in Sections 6 and 7.

DEFINITION 9 (i-trees, **iT**). *An i-tree $\mathcal{T}$ is a simple i-diagram such that $\mathcal{T} = \bot_d$ or such that all of the following three conditions hold:*

1. *$\mathcal{T}$ is **tree shaped***
2. *$\mathcal{T}$ has **independent views***
3. *$\mathcal{T}$ has **independent signatures***.

*We denote by* **iT** *the set of i-trees.*

The following theorem justifies why all three conditions in our definition of i-trees are necessary. Its proof is based on a reduction from graph 3-colorability, which can be encoded using slightly different i-diagrams for each of the three cases. The common property of these diagrams is that they can encode disjointness of arbitrary pairs of nodes.

---
[2] As a result, we can recognize this structure in linear time using, for example, a tree-automaton [12].

THEOREM 2. *Omitting any one out of three conditions from Definition 9 yields a class of diagrams whose satisfiability is NP-hard.*

We note that in addition to NP-hardness, the omission of tree shaped or independent views properties in fact retains the full expressive power of CBAC constraints, using a similar argument as in Lemma 3.

Our ability to specify i-trees as a natural subclass of i-diagrams justifies the definition of i-diagrams themselves. For example, the definition of i-trees would have been more complex had we chosen to represent disjointness using a binary relation $s_1 \cap s_2 = \emptyset$.

Let us also observe that, despite the imposed restrictions, i-trees are fairly expressive. In particular they can express hierarchical decomposition of a set given by a node $S$ into disjoint sets $S_1, \ldots, S_n$, by letting $\{S_1, \ldots, S_n\} \in \mathsf{Split}(S) \cap \mathsf{Comp}(S)$. Despite the independent view condition, we can have multiple orthogonal decompositions, so $\{S'_1, \ldots, S'_m\} \in \mathsf{Split}(S) \cap \mathsf{Comp}(S)$ for $\{S'_1, \ldots, S'_m\} \cap \{S_1, \ldots, S_n\} = \emptyset$. This allows i-trees to naturally express generalized typestate constraints.

## 6. Deciding the Satisfiability of I-Trees

In this section we prove that the satisfiability of i-trees is decidable in polynomial time. For this purpose we introduce a set of *weak consistency* conditions $\mathcal{C}_i$ (Definition 10) such that:

(6.1) We can enforce weak consistency for any satisfiable i-tree using a rewriting system $\mathcal{R}^w$ (Definition 11) with the following properties (Lemma 5):

- $\mathcal{R}^w$ is semantic-preserving;
- if a non-$\bot_d$ i-tree is in $\mathcal{R}^w$ normal form, then it satisfies weak consistency conditions;
- for a particular strategy (Figure 9) the system $\mathcal{R}^w$ terminates in polynomial time.

(6.2) Every i-tree that satisfies weak consistency conditions is satisfiable; Lemma 6 gives an algorithm for constructing a model for any i-tree that satisfies weak consistency conditions.

Figure 9 summarizes the polynomial-time satisfiability decision procedure whose correctness (Theorem 3) follows from the results of this section.

DEFINITION 10 (Weak Consistency). *An i-tree satisfies weak consistency iff $\mathcal{T} \neq \bot_d$ and $\mathcal{T}$ satisfies the following conditions for all $S \in \mathbb{S}$:*

$$\forall S' \in \mathsf{Sons}(S).\ \Phi(S') \supseteq \Phi(S) \qquad (\mathcal{C}_1)$$
$$\mathsf{CSup}(S) > 0 \Rightarrow \forall P \in \mathsf{PN}.\ \{+P, -P\} \not\subseteq \Phi(S) \qquad (\mathcal{C}_2)$$
$$\forall Q \in \mathsf{Comp}(S).\ \mathsf{CSup}(S) \leq \Sigma(\mathsf{CSup}[Q]) \qquad (\mathcal{C}_3)$$
$$\forall Q \in \mathsf{Split}(S).\ \mathsf{CInf}(S) \geq \Sigma(\mathsf{CInf}[Q]) \qquad (\mathcal{C}_4)$$
$$\mathsf{CInf}(S) \leq \mathsf{CSup}(S) \qquad (\mathcal{C}_5)$$

### 6.1 A Rewriting System $\mathcal{R}^w$ for Enforcing Weak Consistency

We introduce the following rewriting system to enforce weak consistency properties when possible.

DEFINITION 11 (System $\mathcal{R}^w$). *For each tuple ($k$,* name, condition, effect*) in Figure 7, we define a rewriting rule on i-diagrams by*

$$\mathcal{D} \xrightarrow[\text{name}]{\text{spot}} \mathcal{D}' \stackrel{def}{\iff} (\mathcal{D} \neq \bot_d \land \text{condition} \land \mathcal{D}' = \mathcal{D}[\text{effect}])$$

*for each assignment* spot *of the free variables appearing in the* condition *column. We define $R_k$ by*

$$\mathcal{D} \xrightarrow[R_k]{} \mathcal{D}' \stackrel{def}{\iff} \exists \text{spot}.\ \mathcal{D} \xrightarrow[\text{name}]{\text{spot}} \mathcal{D}'$$

*We define $\mathcal{R}^w$ as union of $\xrightarrow[R_j]{}$ for $1 \leq j \leq 5$.*

| k | name | conditions | effect |
|---|---|---|---|
| 1 | **DnPhi** | $a_1)$ $S \in \mathsf{Sons}(S')$ <br> $b_1)$ $\phi_n \doteq \Phi(S) \cup \Phi(S')$ <br> $c_1)$ $\phi_n \not\subseteq \Phi(S)$ | $\Phi(S) \leftarrow \phi_n$ |
| 2 | **Unsat** | $a_2)$ $\{+P, -P\} \subseteq \Phi(S)$ <br> $b_2)$ $n \doteq 0$ <br> $c_2)$ $\mathsf{CSup}(S) > n$ | $\mathsf{CSup}(S) \leftarrow n$ |
| 3 | **UpSup** | $a_3)$ $Q \in \mathsf{Comp}(S)$ <br> $b_3)$ $n \doteq \Sigma(\mathsf{CSup}[Q])$ <br> $c_3)$ $\mathsf{CSup}(S) > n$ | $\mathsf{CSup}(S) \leftarrow n$ |
| 4 | **UpInf** | $a_4)$ $Q \in \mathsf{Split}(S)$ <br> $b_4)$ $n \doteq \Sigma(\mathsf{CInf}[Q])$ <br> $c_4)$ $\mathsf{CInf}(S) < n$ | $\mathsf{CInf}(S) \leftarrow n$ |
| 5 | **Error** | $a_5)$ $\mathsf{CInf}(S) > \mathsf{CSup}(S)$ | $\mathcal{D} \leftarrow \bot_d$ |

**Figure 7.** System $\mathcal{R}^w$ for ensuring weak consistency

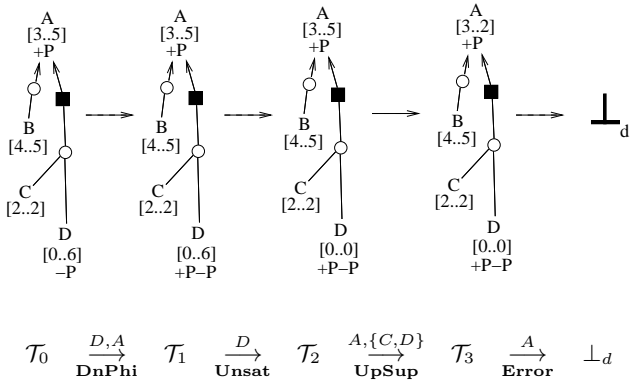

**Figure 8.** An example sequence of rewriting steps for $\mathcal{R}^w$

Figure 8 shows an example sequence of rewriting steps applied to an i-tree.

LEMMA 5 (Properties of $\mathcal{R}^w$).

1. $\mathcal{R}^w$ is **iT**-stable, that is
$$\mathcal{T} \in \mathbf{iT} \wedge \mathcal{T} \xrightarrow[\mathcal{R}^w]{} \mathcal{T}' \Rightarrow \mathcal{T}' \in \mathbf{iT}$$

2. $\mathcal{R}^w$ preserves the semantics, that is
$$\mathcal{D} \xrightarrow[\mathcal{R}^w]{} \mathcal{D}' \Rightarrow \mathcal{D} \equiv \mathcal{D}'$$

3. $\mathcal{R}^w$ enforces weak consistency when possible, that is, a diagram $\mathcal{D}$ in $\mathcal{R}^w$ normal form is either equal to $\bot_d$ or it is weakly consistent

4. $\mathcal{R}^w$ terminates in polynomial time for the strategy corresponding to the algorithm $\mathbf{R}^w_{\mathsf{NF}}$ in Figure 9.

**Proof sketch.**

1. Follows easily from the fact that $\mathcal{R}^w$ rules do not modify $\mathsf{Sons}$, $\mathsf{Split}$, $\mathsf{Comp}$.
2. Follows by construction of $\mathcal{R}^w$ rules. Suppose $\mathcal{D} \xrightarrow[\mathcal{R}^w]{} \mathcal{D}'$. Then $\mathcal{D} \models \mathcal{D}'$ follows from conditions $a_i$ ($1 \leq i \leq 5$), and $\mathcal{D}' \models \mathcal{D}$ follows from conditions $c_i$ ($1 \leq i \leq 4$).
3. For every $k = 1..5$, the condition of application of the rule $R_k$ corresponds to the negation of $\mathcal{C}_k$. When a diagram is in normal form for the rule $R_k$, it either satisfies $\mathcal{C}_k$, or is $\bot_d$.
4. To prove that $\mathbf{R}^w_{\mathsf{NF}}$ corresponds to a polynomial strategy, we prove by induction that applying the rule $R_k$ in the speci-

```
proc R^w_NF(T)
  1. for every S ∈ 𝕊 from the root to the leaves
        for every Q ∈ Comp(S)
            try to apply DnPhi(S, Q) to T
  2. for every S ∈ 𝕊
        for every P ∈ PN
            try to apply Unsat(S, P) to T
  3. for every S ∈ 𝕊 from the leaves to the root
        for every Q ∈ Comp(S)
            try to apply UpSup(S, Q) to T
  4. for every S ∈ 𝕊 from the leaves to the root
        for every Q ∈ Split(S)
            try to apply UpInf(S, Q) to T
  5. for every S ∈ 𝕊
        try to apply Error(S) to T
  return T

proc ItreeSAT(T)
  if (R^w_NF(T) = ⊥_d) return satisfiable
  else return unsatisfiable
```

**Figure 9.** Polynomial-time algorithms $\mathbf{R}^w_{\mathsf{NF}}$ and **ItreeSAT** to compute $\mathcal{R}^w$ normal form and check satisfiability of i-trees

fied direction (from the root to the leaves or from the leaves to the root), enforces $\mathcal{C}_k$ everywhere, and when $\mathcal{C}_k$ holds, the rule is not applicable anymore. Finally, we prove that each rule $R_k$ for $k = 1..5$ preserves the conjunction of properties $\bigwedge_{j=1..(k-1)} R_k$, and as a consequence, we never need to reapply any of the rules $R_j$ for $j < k$. ∎

### 6.2 Constructing Models for Weakly Consistent I-Trees

The following Lemma 6 is crucial for the completeness of our algorithm, and justifies the definition of weak consistency.

LEMMA 6 (Model Construction). *If an i-tree $\mathcal{T}$ is weakly consistent, then we can construct a model for $\mathcal{T}$.*

The high-level idea of the proof of Lemma 6 is to first build the first two components $(\Delta, \alpha)$ of the model, and then extend the model with $\Xi$ using the independent signatures condition for i-trees. We build the $(\Delta, \alpha)$ part of the model by building a model for each subtree using an induction on the height of the i-tree. To construct models that satisfy $\mathsf{Split}$ and $\mathsf{Comp}$ constraints in the inductive step, we use a stronger induction hypothesis: we show that there exists a model $(\Delta, \alpha)$ for a tree rooted in node $S$ with $|\Delta| = k$ for all $\mathsf{CInf}(S) \leq k \leq \mathsf{CSup}(S)$, and we rely on the properties of weak consistency to prove the inductive step. The proof of this lemma is interesting because similar ideas are used when building example models that show the completeness in Section 7.

Putting all results in this section together using the argument at the beginning of the section, we obtain the following theorem.

THEOREM 3 (**ItreeSAT** Correctness). *$\mathcal{T}$ is satisfiable if and only if $\mathbf{R}^w_{\mathsf{NF}}(\mathcal{T}) \neq \bot_d$. Therefore, the algorithm **ItreeSAT** in Figure 9 is a sound and complete polynomial-time decision procedure for the satisfiability of i-trees.*

## 7. Deciding Subsumption of I-Trees

The goal of this section is to prove that we can decide the subsumption of i-trees in polynomial time. Note that the subclass of i-trees is not closed under negation or implication, so we cannot decide $\mathcal{T} \models \mathcal{T}'$ by checking the satisfiability of $\neg(\mathcal{T} \Rightarrow \mathcal{T}')$. Instead, our approach is to bring $\mathcal{T}$ into a form where the properties of the models of $\mathcal{T}$ are *easy to read* from $\mathcal{T}$. We then check that $\mathcal{T}$ entails each of the conditions that correspond to the semantics of $\mathcal{T}'$.

| $k$ | name | condition | effect |
|---|---|---|---|
| 6 | **DnInf** | $a_6)$ $(\{S\} \uplus Q_0) \in \mathsf{Comp}(S')$ <br> $b_6)$ $n \doteq \mathsf{CInf}(S')$ <br> $\quad\quad -\Sigma(\mathsf{CSup}[Q_0])$ <br> $c_6)$ $n > \mathsf{CInf}(S)$ | $\mathsf{CInf}(S) \leftarrow n$ |
| 7 | **DnSup** | $a_7)$ $(\{S\} \uplus Q_0) \in \mathsf{Split}(S')$ <br> $b_7)$ $n \doteq \mathsf{CSup}(S')$ <br> $\quad\quad -\Sigma(\mathsf{CInf}[Q_0])$ <br> $c_7)$ $n < \mathsf{CSup}(S)$ | $\mathsf{CSup}(S) \leftarrow n$ |
| 8 | **CCmp**[*] | $a_8)$ $Q \in \mathsf{Split}(S) \wedge$ <br> $\quad\quad \mathsf{CSup}(S) \leq \Sigma(\mathsf{CInf}[Q])$ <br> $b_8)$ $\mathbb{C}_n \doteq \mathsf{Comp}(S) \cup \{Q\}$ <br> $c_8)$ $\mathbb{C}_n \not\subseteq \mathsf{Comp}(S)$ | $\mathsf{Comp}(S) \leftarrow \mathbb{C}_n$ |
| 9 | **CSplit**[*] | $a_9)$ $Q \in \mathsf{Comp}(S) \wedge$ <br> $\quad\quad \mathsf{CInf}(S) \geq \Sigma(\mathsf{CSup}[Q])$ <br> $b_9)$ $\mathbb{C}_n \doteq \mathsf{Split}(S) \cup \{Q\}$ <br> $c_9)$ $\mathbb{C}_n \not\subseteq \mathsf{Split}(S)$ | $\mathsf{Split}(S) \leftarrow \mathbb{C}_n$ |
| 10 | **UpPhi** | $a_{10})$ $Q \in \mathsf{Comp}(S)$ <br> $b_{10})$ $\phi_n \doteq \Phi(S) \cup \bigcap \Phi[Q]$ <br> $c_{10})$ $\phi_n \not\subseteq \Phi(S)$ | $\Phi(S) \leftarrow \phi_n$ |
| 11 | **Void**[*] | $a_{11})$ $S \neq \emptyset_d \wedge \mathsf{CSup}(S) = 0$ | $\mathsf{Merge}(\{S, \emptyset_d\})$ |
| 12 | **Equal**[*] | $a_{12})$ $\{S'\} \in \mathsf{Comp}(S)$ | $\mathsf{Merge}(\{S, S'\})$ |

[*]Follow the application of these rules by **Simplify**.

**Figure 10.** Rules for System $\mathcal{R}$

We formalize the intuitive condition of being easy to read in the notion of *strong consistency*. We build on the system $\mathcal{R}^w$ from the previous section to create a larger rewriting system $\mathcal{R}$ for ensuring strong consistency. We introduce a polynomial-time strategy for $\mathcal{R}$ that transforms every i-tree into $\bot_d$ or into an i-tree that is strongly consistent, and we give polynomial-time algorithms for extracting the information from strongly consistent i-trees.

DEFINITION 12 (Strong Consistency). *An i-tree $\mathcal{T}$ is* strongly consistent *iff it is weakly consistent and satisfies all of the following properties:*

$\forall Q \in \mathsf{Comp}(S).\ \forall S_0 \in Q.$
$\quad \mathsf{CInf}(S_0) \geq \mathsf{CInf}(S) - \Sigma(\mathsf{CSup}[Q - \{S_0\}])$ $\quad (\mathcal{C}_6)$

$\forall Q \in \mathsf{Split}(S).\ \forall S_0 \in Q.$
$\quad \mathsf{CSup}(S_0) \leq \mathsf{CSup}(S) - \Sigma(\mathsf{CInf}[Q - \{S_0\}])$ $\quad (\mathcal{C}_7)$

$\forall Q \in \mathsf{Split}(S).\ Q \notin \mathsf{Comp}(S) \Rightarrow$
$\quad \mathsf{CSup}(S) > \Sigma(\mathsf{CInf}[Q])$ $\quad (\mathcal{C}_8)$

$\forall Q \in \mathsf{Comp}(S).\ Q \notin \mathsf{Split}(S) \Rightarrow$
$\quad \mathsf{CInf}(S) < \Sigma(\mathsf{CSup}[Q])$ $\quad (\mathcal{C}_9)$

$\forall Q \in \mathsf{Comp}(S).\ \bigcap(\Phi[Q]) \subseteq \Phi(S)$ $\quad (\mathcal{C}_{10})$

$S \neq \emptyset_{\mathcal{D}} \Rightarrow \mathsf{CSup}(S) > 0$ $\quad (\mathcal{C}_{11})$

$Q \in \mathsf{Comp}(S) \Rightarrow |Q| > 1$ $\quad (\mathcal{C}_{12})$

### 7.1 A rewriting system $\mathcal{R}$ to enforce strong consistency

This section follows the development of Section 6.1.

DEFINITION 13 (System $\mathcal{R}$). *The system $\mathcal{R}$ extends $\mathcal{R}^w$ with the additional rules of Figure 10, analogously to Definition 11.*

LEMMA 7 (Properties of $\mathcal{R}$). *1. $\mathcal{R}$ is **iT**-stable, that is*

$$\mathcal{T} \in \mathbf{iT} \wedge \mathcal{T} \xrightarrow[\mathcal{R}]{} \mathcal{T}' \Rightarrow \mathcal{T}' \in \mathbf{iT}$$

*2. $\mathcal{R}$ preserves the semantics, that is*

$$\mathcal{D} \xrightarrow[\mathcal{R}]{} \mathcal{D}' \Rightarrow \mathcal{D} \equiv \mathcal{D}'$$

```
proc R_NF(T)
    1...5. T ← R_NF^w(T)
    6. for each S' ∈ 𝕊 from the root to the leaves
           for each Q ∈ Comp(S)
               for every S ∈ Q
                   try DnInf(S, S', Q)
    7. for each S' ∈ 𝕊 from the root to the leaves
           for each Q ∈ Split(S)
               for each S ∈ Q
                   try DnSup(S, S', Q)
    8. for each S ∈ 𝕊
           for each Q ∈ Split(S)
               try CCmp(S, Q)
    9. for each S ∈ 𝕊
           for each Q ∈ Comp(S)
               try CSplit(S, Q)
   10. for each S ∈ 𝕊 from the leaves to the root
           for each Q ∈ Comp(S)
               try UpPhi(S, Q)
   11. for each S ∈ 𝕊 from the leaves to the root
               try Void(S)
   12. for each S ∈ 𝕊
           for each Q ∈ Comp(S)
               try Equal(S, Q)
    return T
```

**Figure 11.** Polynomial-time algorithm $\mathbf{R}_{\mathsf{NF}}(\mathcal{T})$ to compute $\mathcal{R}$ normal form

3. $\mathcal{R}$ *enforces strong consistency when possible, that is, a diagram $\mathcal{D}$ in $\mathcal{R}$ normal form is either equal to $\bot_d$ or it is strongly consistent.*

4. $\mathcal{R}$ *terminates in polynomial time for the strategy corresponding to the algorithm $\mathbf{R}_{\mathsf{NF}}$ described in Figure 11.*

**Proof sketch.**

1. The **iT**-stability is trivial for the rules **DnInf**, **DnSup**, **UpPhi**. The other rules are marked with a star and we use the algorithm Simplify. In fact, we can show that it is not necessary to apply **Simplify** in its full generality, but only to remove any redundant views introduced by **CCmp** and **CSplit**, remove any self edges introduced by the operation Merge used in the rules **Equal** and **Void**, and to remove the edges going to $\emptyset_d$ that can be introduced by the rule **Void**.

2,3. Follow by construction as in the previous section.

4. This part is significantly more difficult than for system $\mathcal{R}^w$, because the interactions between the rules are more complex, but follows the same structure as the proof for $\mathcal{R}^w$.

### 7.2 Extracting Information from Strongly Consistent I-Trees

In this section we start from a strongly consistent i-tree $\mathcal{T}$ and consider the problem of checking $\mathcal{T} \models \mathcal{D}'$. Analyzing Definition 2, we observe that a diagram corresponds to a conjunction of constraints. Therefore, the subsumption problem $\mathcal{T} \models \mathcal{D}'$ corresponds to the problem of verifying that $\mathcal{T}$ entails atomic formulas of the form $s = \emptyset$, $s_1 = s_2$, $s_1 \subseteq s_2$, $a \leq |s| \leq b$, $s \subseteq P$, $s \subseteq P^c$, $s_1 \cap s_2 = \emptyset$ and $s \subseteq \bigcup\{s_1, \ldots, s_n\}$. Without the danger of confusion, we write $\mathcal{T} \models A$ when the atomic formula $A$ holds in all models for $\mathcal{T}$.

THEOREM 4. *Let $\mathcal{T}$ be a strongly consistent i-tree and let $\mathsf{H}_A^{\mathcal{T}}$ for atomic formula $A$ be as defined in Figure 12. Then $\mathcal{T} \models A$ if and only if $\mathsf{H}_A^{\mathcal{T}}$.*

proc **Subsumes**$(\mathcal{T}, \mathcal{D}')$
  $\mathcal{T} := \mathbf{R}_{\mathsf{NF}}(\mathcal{T})$
  let $f : \mathsf{SN} \to \mathbb{S}$ such that $\forall s \in \mathsf{SN}.\ s \in f(s)$
  let $h' : \mathbb{S}' \to \mathsf{SN}$ be any function such that $\forall S' \in \mathbb{S}.\ h'(S') \in S'$
  check all of the following conditions:
  1. $\bigwedge_{S \in \mathbb{S}'} \bigwedge_{s_1, s_2 \in S} \mathsf{H}^{\mathcal{T}}_{s_1 = s_2}$
     where $\mathsf{H}^{\mathcal{T}}_{s_1 = s_2} \stackrel{def}{\iff} f(s_1) = f(s_2)$
  2. $\mathsf{H}^{\mathcal{T}}_{h(\emptyset'_d) = \emptyset}$
     where $\mathsf{H}^{\mathcal{T}}_{s = \emptyset} \stackrel{def}{\iff} f(s) = \emptyset_d$
  3. $\bigwedge_{S \in \mathbb{S}'} \mathsf{H}^{\mathcal{T}}_{\mathsf{CInf}'(S) \leq |h(S)| \leq \mathsf{CSup}'(S)}$
     where $\mathsf{H}^{\mathcal{T}}_{a \leq |s| \leq b} \stackrel{def}{\iff} \mathsf{CInf}(f(s)) \leq a \leq b \leq \mathsf{CSup}(f(s))$
  4. $\bigwedge_{S \in \mathbb{S}'} \bigwedge_{(+P) \in \Phi'(S)} \mathsf{H}^{\mathcal{T}}_{+P(h(S))}$
     where $\mathsf{H}^{\mathcal{T}}_{+P(s)} \stackrel{def}{\iff} (+P) \in \Phi(f(s))$
  5. $\bigwedge_{S \in \mathbb{S}'} \bigwedge_{(-P) \in \Phi'(S)} \mathsf{H}^{\mathcal{T}}_{-P(h(S))}$
     where $\mathsf{H}^{\mathcal{T}}_{-P(s)} \stackrel{def}{\iff} (-P) \in \Phi(f(s))$
  6. $\bigwedge_{S \in \mathbb{S}'} \bigwedge_{S' \in \mathsf{Sons}'(S)} \mathsf{H}^{\mathcal{T}}_{h(S) \subseteq h(S')}$
     where $\mathsf{H}^{\mathcal{T}}_{s_1 \subseteq s_2} \stackrel{def}{\iff} f(s_1) = \emptyset_d \lor f(s_1) \stackrel{*}{\leadsto} f(s_2)$
  7. $\bigwedge_{S \in \mathbb{S}'} \bigwedge_{Q \in \mathsf{Split}'(S)} \bigwedge_{\substack{S_1, S_2 \in Q \\ S_1 \neq S_2}} \mathsf{H}^{\mathcal{T}}_{h(S_1) \cap h(S_2) = \emptyset}$
     where $\mathsf{H}^{\mathcal{T}}_{s_1 \cap s_2 = \emptyset} \stackrel{def}{\iff} f(s_1) = \emptyset_d \lor f(s_2) = \emptyset_d \lor \mathsf{disj}^*_{\mathcal{T}}(S_1, S_2)$
  8. $\bigwedge_{S \in \mathbb{S}'} \bigwedge_{Q \in \mathsf{Comp}'(S)} \mathsf{H}^{\mathcal{T}}_{h(S) \subseteq \cup h[Q]}$
     where $\mathsf{H}^{\mathcal{T}}_{s \subseteq \cup Z} \stackrel{def}{\iff} f(s) = \emptyset_d \lor \mathbf{Included}(f(s), f[Z], \mathcal{T})$
     where proc **Included**$(S_0, C, \mathcal{T})$
       return $\bigvee_{S_0 \stackrel{*}{\leadsto} S} \mathbf{Incl}(S)$
     proc **Incl**$(S)$
       if $S \in C$ then return **true**
       else return $\bigvee_{Q \in \mathsf{Comp}(S)} \left( \bigwedge_{S' \in Q} \mathbf{Incl}(S') \right)$

**Figure 12.** An Algorithm for Computing $\mathcal{T} \models \mathcal{D}'$ for a an i-tree $\mathcal{T}$ and an arbitrary diagram $\mathcal{D}'$.

It is easy to verify that $\mathsf{H}^{\mathcal{T}}_A$ implies $\mathcal{T} \models A$. The proof of the converse is based on the following two lemmas, which provide a link between strong and weak consistency.

LEMMA 8 (Bounds Refinement). *Let $\mathcal{T}$ be a strongly consistent i-tree, $S \in \mathbb{S}$, $i, s$ such that $\mathsf{CInf}(S) \leq i \leq s \leq \mathsf{CSup}(S)$, let $\mathcal{T}' = \mathcal{T}[\mathsf{CInf}(S) \leftarrow i, \mathsf{CSup}(S) \leftarrow s]$ and $\mathcal{T}'_{\mathsf{NF}} = \mathbf{R}^w_{\mathsf{NF}}(\mathcal{T}')$. Then 1) $\mathcal{T}'_{\mathsf{NF}} \neq \bot_d$, 2) $\mathcal{T}'_{\mathsf{NF}} \models \mathcal{T}$, and 3) if $\neg(S \stackrel{*}{\leadsto} S_0)$, then*

$$(\mathsf{CInf}(S_0), \mathsf{CSup}(S_0)) = (\mathsf{CInf}'_{\mathsf{NF}}(S_0), \mathsf{CSup}'_{\mathsf{NF}}(S_0)).$$

***Proof sketch.***

1. We prove this result by induction on the depth of $S$ in the tree $(\mathbb{S}, \leadsto)$. The key step of this proof is to show that the application of UpSup and/or UpInf to the father $S'$ of $S$ does not produce a situation where $a_5$ holds in the resulting diagram $\mathcal{T}''$ (and therefore the rule Error is not applicable in $\mathcal{T}''$). We use the fact that $\mathbf{R}^w_{\mathsf{NF}}$ applies the rules UpInf and UpSup bottom up, and prove that each application preserves $\mathcal{C}_5$, only increases CInf and only decreases CSup. At each step we distinguish three cases:
   (a) both UpSup and UpInf are applicable; then the result follows from $\mathcal{C}_5$;
   (b) only UpSup is applicable; then the result follows from $\mathcal{C}_6$;
   (c) only UpInf is applicable; then the result follows from $\mathcal{C}_7$.

2. Follows easily from the hypothesis $\mathsf{CInf}(S) \leq i \leq s \leq \mathsf{CSup}(S)$ and the fact that $\mathbf{R}^w_{\mathsf{NF}}$ is semantics preserving.

3. It is enough to notice that only rules UpInf and UpSup are used when applying $\mathbf{R}^w_{\mathsf{NF}}$, and these rules are applied in the bottom-up direction. ∎

The fact that the resulting i-tree $\mathcal{T}'_{\mathsf{NF}}$ is not strongly consistent anymore prevents us to apply this lemma twice from a given strongly consistent i-tree. To enforce more than one restriction, we need to refine simultaneously the bounds of several nodes. For this purpose, we use the following lemma.

LEMMA 9 (Parallel Bounds Refinement). *Let $\mathcal{T}$ be a strongly consistent i-tree, and $(Q_0, \leadsto)$ a subtree of $\mathcal{T}$ such that*

- *The nodes of $Q_0$ are pairwise independent, that is, $\forall S_1, S_2 \in Q_0.\ \neg(\mathsf{disj}^*_{\mathcal{T}}(S_1, S_2))$*
- *$(Q_0, \leadsto)$ has the same root as $\mathcal{T}$.*

*Then the i-tree $T'$ defined by the simultaneous update*

$$\mathcal{T}' \stackrel{def}{=} \mathcal{T}[\ \forall S \in Q_0 : \mathsf{CInf}(S) \leftarrow \mathsf{CSup}(S)\ ]$$

*is such that its $\mathcal{R}^w$ normal form $\mathcal{T}'_{\mathsf{NF}} \stackrel{def}{=} \mathbf{R}^w_{\mathsf{NF}}(\mathcal{T}')$ satisfies*

1. $\mathcal{T}'_{\mathsf{NF}} \neq \bot_d$
2. $\mathcal{T}'_{\mathsf{NF}} \models \mathcal{T}$

Lemmas 8 and 9 are the basic tools we need to show that the information syntactically computed from an i-tree is the most precise information computable from the semantics of the i-tree. We prove this property for each of the atomic formulas $A$.

LEMMA 10. *If an i-tree $\mathcal{T}$ is strongly consistent, then for all $S \in \mathbb{S}$ we have*

$$S \neq \emptyset_d \Rightarrow \exists \mathcal{M}.\ \mathcal{M} \models \mathcal{T} \land \overline{\alpha}_{\mathcal{M}}(S) \neq \emptyset$$

***Proof.*** If $S \neq \emptyset_d$, we have $\mathsf{CSup}(S) > 0$ by $\mathcal{C}_{11}$ and therefore the i-tree $\mathcal{T}' \stackrel{def}{=} \mathcal{T}[\mathsf{CInf}(S) \leftarrow \max(1, \mathsf{CInf}(S))]$ subsumes $\mathcal{T}$. By Lemma 8, $\mathcal{T}'$ is satisfiable, and we can take any model of $\mathcal{T}'$ as a model of $\mathcal{T}$. ∎

LEMMA 11. *If an i-tree $\mathcal{T}$ is strongly consistent, then for all $S \in \mathbb{S}$ we have*

$$\exists \mathcal{M}.\ \mathcal{M} \models \mathcal{T} \land |\overline{\alpha}_{\mathcal{M}}(S)| = \mathsf{CInf}(S)$$
$$\exists \mathcal{M}.\ \mathcal{M} \models \mathcal{T} \land |\overline{\alpha}_{\mathcal{M}}(S)| = \mathsf{CSup}(S)$$

***Proof.*** According to Lemma 8, the two i-trees $\mathcal{T}'_1 \stackrel{def}{=} \mathcal{T}[\mathsf{CSup}(S) \leftarrow \mathsf{CInf}(S)]$ and $\mathcal{T}'_2 \stackrel{def}{=} \mathcal{T}[\mathsf{CInf}(S) \leftarrow \mathsf{CSup}(S)]$ are satisfiable, and both $\mathcal{T}'_1$ and $\mathcal{T}'_2$ trivially subsume $\mathcal{T}$. Any model $\mathcal{M}_1$ of $\mathcal{T}'_1$ is such that $|\overline{\alpha}_{\mathcal{M}_1}(S)| = \mathsf{CInf}(S)$, and any model $\mathcal{M}_2$ of $\mathcal{T}'_2$ is such that $|\overline{\alpha}_{\mathcal{M}_2}(S)| = \mathsf{CSup}(S)$. ∎

LEMMA 12. *If an i-tree $\mathcal{T}$ is strongly consistent, $S_0 \in \mathbb{S}$, $C \in \mathcal{P}(\mathbb{S})$, and **Included**$(S_0, C, \mathcal{T})$ returns* false*, then*

$$\exists \mathcal{M}.\ \mathcal{M} \models \mathcal{T} \land \overline{\alpha}_{\mathcal{M}}(S_0) \not\subseteq \bigcup \overline{\alpha}_{\mathcal{M}}[C]$$

***Proof sketch.*** Assume that **Included** returns *false*. We argue that the model $\mathcal{M}$ exists in several steps. Let $Q_0$ be the smallest

set of nodes such that:
$$S_0 \stackrel{*}{\leadsto} S \Rightarrow S \in Q_0$$
$$\left.\begin{array}{r}S \in Q_0 \wedge C_1 \in \mathsf{Comp}(S) \\ \wedge\; S_1 \in C_1 \wedge \neg\mathbf{Incl}(S_1)\end{array}\right\} \Rightarrow S_1 \in Q_0$$

By definition of **Incl**, we have $Q_0 \cap C = \emptyset$.

$Q_0$ is tree-shaped by construction, but may contain two nodes which are explicitly disjoint. We therefore compute a subtree $Q_1$ of $Q_0$, by starting from the root and keeping at most one son for each complete view. In this process we ensure that $Q_1$ contains $S_0$, by avoiding to cut the branch which leads to $S_0$.

We then apply Lemma 9 to $Q_1$ and construct a model of the resulting i-tree while enforcing that a certain element $x \in \overline{\alpha}(S)$ is such that $x \in \overline{\alpha}(S') \Leftrightarrow S' \in Q_1$ for all $S' \in \mathbb{S}$. More precisely, we prove by induction on $n$ that for each node $S_1$ of $Q_1$ of depth $n$ in the tree $Q_1$ we can construct a model $(\Delta, \alpha, \Xi)$ for the sub-i-tree of $\mathcal{T}$ with root $S_1$ such that
$$\forall S'.\; S' \stackrel{*}{\leadsto} S_1 \Rightarrow (x \in \overline{\alpha}(S') \Leftrightarrow S' \in Q_1)$$

If $n = 0$ then $S_1$ is a leaf of $(Q_1, \leadsto)$ and has no complete view by construction of $Q_1$. Then using $\mathcal{C}_8$ we show that we can construct a model of the sub-i-tree with root $S_1$ containing a fresh element (not included in any of the sons of $S_1$).

If $n > 0$, we can deal with the split views in the same way, but this time $S_1$ can have some complete views. If this complete view contains a unique split view, we avoid merging $x$ in a son of $S_1$ with elements in the other sons of $S_1$. If there exist more than one split view, we can use $\mathcal{C}_8$ and construct the model using a refinement of the ideas of Lemma 6.

Finally, since $S_0 \in Q_1$, we have $x \in \overline{\alpha}(S_0)$, and since $C \cap Q_1 = \emptyset$ we have $x \notin \bigcup \overline{\alpha}[C]$. ∎

LEMMA 13. *If an i-tree $\mathcal{T}$ is strongly consistent, for all $S_1, S_2 \in \mathbb{S}$ such that $S_1 \neq \emptyset$ and $S_2 \neq \emptyset$ we have*
$$\neg(S_1 \stackrel{*}{\leadsto} S_2) \Rightarrow \exists \mathcal{M}.\; \mathcal{M} \models \mathcal{T} \wedge \overline{\alpha}_\mathcal{M}(S_1) \not\subseteq \overline{\alpha}_\mathcal{M}(S_2)$$

**Proof.** The property $\exists \mathcal{M}.\; \mathcal{M} \models \mathcal{T} \wedge \overline{\alpha}_\mathcal{M}(S_1) \not\subseteq \overline{\alpha}_\mathcal{M}(S_2)$ can be checked using $\mathbf{Included}(S_1, \{S_2\}, \mathcal{T})$. Using $\mathcal{C}_{12}$ we show that this test is equivalent to test $S_1 \stackrel{*}{\leadsto} S_2$. ∎

LEMMA 14. *If an i-tree $\mathcal{T}$ is strongly consistent, for all $S_1, S_2 \in \mathbb{S}$ we have*
$$S_1 \neq S_2 \Rightarrow \exists \mathcal{M}.\; \mathcal{M} \models \mathcal{T} \wedge \overline{\alpha}_\mathcal{M}(S_1) \neq \overline{\alpha}_\mathcal{M}(S_2)$$

**Proof.** If $S_1 \neq S_2$, then $\neg(S_1 \stackrel{*}{\leadsto} S_2)$ or $\neg(S_2 \stackrel{*}{\leadsto} S_1)$. In either case the result follows from Lemma 13. ∎

LEMMA 15. *If an i-tree $\mathcal{T}$ is strongly consistent, then for all $S \in \mathbb{S}$ and $P \in \mathsf{PN}$ we have*
$$(+P) \notin \Phi(S) \Rightarrow \exists \mathcal{M}.\; \mathcal{M} \models \mathcal{T} \wedge \overline{\alpha}_\mathcal{M}(S) \not\subseteq \Xi(P)$$
$$(-P) \notin \Phi(S) \Rightarrow \exists \mathcal{M}.\; \mathcal{M} \models \mathcal{T} \wedge \overline{\alpha}_\mathcal{M}(S) \not\subseteq \Xi(P)^c$$

**Proof sketch.** Let $S \in \mathbb{S}$ be such that $(+P) \notin \Phi(S)$. We define $Q_P \stackrel{def}{=} \{S' \in \mathbb{S} | (+P) \in \Phi(S')\}$. Using $\mathcal{C}_{10}$ and $\mathcal{C}_1$ we show that $\mathbf{Included}(S, Q_P, \mathcal{T})$ returns false. By Lemma 12, there exists a model such that $\overline{\alpha}(S) \not\subseteq \bigcup \overline{\alpha}[Q_P]$. We then change the model by redefining $\Xi'$ on PN as $\Xi'(P) = \bigcup \overline{\alpha}[Q_P]$, so $\overline{\alpha}(S) \not\subseteq \Xi'(P)$. The case $(-P) \notin \Phi(S)$ is dual and follows from the previous case by swapping $(+P)$ and $(-P)$ in the i-tree and taking complements of $\Xi(P)$. ∎

LEMMA 16. *If an i-tree $\mathcal{T}$ is strongly consistent, then for all $S_1, S_2 \in \mathbb{S}$ such that $S_1 \neq \emptyset_d$, $S_2 \neq \emptyset_d$ we have*
$$\neg(\mathsf{disj}_\mathcal{T}^*(S_1, S_2)) \Rightarrow \exists \mathcal{M}.\; \mathcal{M} \models \mathcal{T} \wedge \overline{\alpha}_\mathcal{M}(S_1) \cap \overline{\alpha}_\mathcal{M}(S_2) \neq \emptyset$$

From the previously stated lemmas, we can prove Theorem 4. From Theorem 4 and Lemma 7 we conclude that the algorithm in Figure 12 is a correct and complete test for subsumption, not only of between trees, but also between a tree and an arbitrary diagram.

## 8. Related Work

***Boolean algebras with cardinalities.*** Quantifier-free formulas of boolean algebra are NP-complete [33]. Quantified formulas of boolean algebra are in alternating exponential space with a linear number of alternations [21]. Cardinality constraints naturally arise in quantifier elimination for boolean algebras [31, 42, 43]. Quantifier elimination implies that each first-order formula of the language of boolean algebras is equivalent to some quantifier-free formula with constant cardinalities; however, quantifier elimination may introduce an exponential blowup. The first-order theory of boolean algebras of finite sets with symbolic cardinalities, or, equivalently, boolean algebras of sets with equicardinality operator is shown decidable in [14]. These results are repeated, motivated by constraint solving applications, in [23, 39] and a special case with quantification over elements only is presented in [44]. Upper and lower bounds on the complexity of this problem were shown in [22] which also introduces the name BAPA, for Boolean Algebra with Presburger Arithmetic. The quantifier-free case of BAPA was studied in [45] with an NEXPTIME decision procedure, which is also achieved as a special case of [23, 22]. The new decision procedure in the present paper improves this bound to PSPACE and gives insight into the problem by reducing it to boolean algebras with binary-encoded large cardinalities, and showing that it is not necessary to explicitly construct all set partitions.

Several decidable fragments of set theory are studied in [10]. Cardinality constraints also occur in description logics [5] and two-variable logic with counting [35, 19, 38]. However, all logics of counting that we are aware of have complexity that is beyond PSPACE.

We are not aware of any previously known fragments of boolean algebras of sets with cardinality constraints that have polynomial-time satisfiability or subsumption algorithms. Our polynomial-time result for i-trees is even more interesting in the light of the fact that our constraints can express some "disjunction-like" properties such as $A = B \cup C$.

***Set constraints.*** Set constraints [1, 3, 2, 6] are incomparable to the constraints studied in our paper. On the one hand, set constraints are interpreted over ground terms and contain operations that apply a given free function symbol to each element of the set, which makes them suitable for encoding type inference [4] and interprocedural analysis [20, 34]. Researchers have also explored the efficient computation of the subset relation for set constraints [13]. On the other hand, set constraints do not support cardinality operators that are useful in modelling databases [40, 11] and analysis of the sizes of data structures [29]. Tarskian constraints use uninterpreted function symbols instead of free function symbols and have very high complexity [18].

## 9. Conclusions

Constraints on sets and relations are very useful for analysis of software artifacts and their abstractions. Reasoning about sets and relations often involves reasoning about their sizes. For example, an integer field may be used to track the size of the set of objects stored in a data structure. In this paper, we have presented new complexity results and algorithms for solving constraints on boolean algebra of sets with symbolic and constant cardinality constraints. We have presented symbolic constraints and large constant constraints, gave more efficient algorithm for quantifier-free symbolic constraints, identified several sources of NP-hardness of constraints, and presented a new class of constraints for which satisfiability and en-tailment are solvable in polynomial time. We hope that our results will serve as concrete recipes and general guidance in the design of algorithms for constraint solving and program analysis.

## A. Proofs

### A.1 I-Diagrams

**Lemma 3** *I-diagrams have the same expressive power as* CBAC *constraints.*

*Proof.* We translate an i-diagram into a CBAC constraint as follows. As in Figure 2, we note that $b_1 = b_2$, $b_1 \subseteq b_2$ can be expressed in the form $|b| = 0$, so we may assume that they are part of CBAC. Similarly, $|b| \leq k$ can be expressed as $b \subseteq s \land |s| = k$ for a fresh variable $s$, and $|b| \geq k$ can be expressed as $s \subseteq b \land |s| = k$. We translate $\perp_d$ into e.g. $|\mathbf{0}| = 1$. Next consider $\mathcal{D} = (\mathbb{S}, \emptyset_d, \mathsf{Sons}, \mathsf{Split}, \mathsf{Comp}, \mathsf{CInf}, \mathsf{CSup}, \Phi)$. For each $S \in \mathbb{S}$, let $\eta(S) \in S$ be a representative set name. For each $S_1 \in S \setminus \{\eta(S)\}$ introduce conjunct $S_1 = \eta(S)$. Next, for each $S_1 \in \mathsf{Sons}(S)$, introduce a conjunct $S_1 \subseteq S$. For each $+P \in \Phi(S)$, introduce conjunct $S \subseteq P$, and for each $-P \in \Phi(S)$ conjunct $S \subseteq P^c$. Express the bounds using conjuncts $|S| \leq \mathsf{CSup}(S)$ and $|S| \geq \mathsf{CInf}(S)$. For each $Q \in \mathsf{Split}(S)$ and $S_1, S_2 \in Q$ where $S_1 \neq S_2$, introduce conjunct $|S_1 \cap S_2| = 0$. For each $\{S_1, \ldots, S_n\} \in \mathsf{Comp}(S)$ introduce conjunct $S = S_1 \cup \ldots \cup S_n$.

We translate a CBAC constraint into i-diagram using the following observations. It is sufficient to translate the following boolean algebra expressions: $s_0 = s_1 \cup s_2$, $s_0 = s_1^c$, and $|s| = k$. We construct an i-diagram whose nodes are singletons. We pick one set variable $u$ to act as a universal set and put $\{s\} \in \mathsf{Sons}(\{u\})$ for every set variable $s$ in the i-diagram. We translate $s_0 = s_1 \cup s_2$ as $\{\{s_1\}, \{s_2\}\} \in \mathsf{Comp}(\{s_0\})$ and translate $s_0 = s_1^c$ as $\{\{s_0\}, \{s_1\}\} \in \mathsf{Split}(\{u\})$, $\{\{s_0\}, \{s_1\}\} \in \mathsf{Comp}(\{u\})$. We translate $|s| = k$ as $\mathsf{CInf}(\{s\}) = k$ and $\mathsf{CSup}(\{s\}) = k$. Then for each satisfiable assignment of CBAC there is a model for the constructed i-diagram where $\{u\}$ is interpreted as a universal set. Conversely, for a model of i-diagram where $\bar{\alpha}(\{s\}) = A$, we let $[s \mapsto A \cap \bar{\alpha}(u)]$. The result is an assignment that satisfies the original CBAC formula. ∎

**Lemma 4** *For every i-diagram $\mathcal{D}$ we can obtain an equivalent simple i-diagram using the polynomial-time algorithm in Figure 6.*

*Proof.* We argue that algorithm in Figure 6 produces diagram that is *i)* well-formed, *ii)* simple, *iii)* equivalent to the original diagram.

We first observe that after step 2, the following two conditions hold:

C) if $Q \in \mathsf{Split}(S)$ and $S \in Q$, then $Q \subseteq \{S, \emptyset_d\}$. This condition holds because the step 1.5 of the algorithm merges all nodes $Q \setminus \{S\}$ with $\emptyset_d$ when $S \in Q \in \mathsf{Split}(S)$.

D) if $Q \in \mathsf{Comp}(S)$, then $Q \neq \emptyset$ (by step 1.3), if $Q = \{\emptyset_d\}$ then $S = \emptyset_d$ (by step 1.6).

*i)* To see that the resulting i-diagram is well-formed, it suffices to check the conditions $\bigcup \mathsf{Split}(S) = \mathsf{Sons}(S)$ and $\bigcup \mathsf{Comp}(S) \subseteq \mathsf{Sons}(S)$. This condition is preserved by factor-diagram construction (for any equivalence relation). It is preserved by step 3 for the following reason. The only nodes removed from $\mathsf{Sons}(S)$ are $\emptyset_d$ and $S$. These nodes do not appear in $\bigcup \mathsf{Comp}(S)$ because $\emptyset_d$ is removed from each view $Q$, and views with $S \in Q$ are removed. It remains to check that $\mathsf{Sons}(S) \subseteq \bigcup \mathsf{Split}(S)$ after step 3, and this holds because our condition (C) implies that no element other than $S, \emptyset_d$ is lost from $\bigcup \mathsf{Split}(S)$ in step 3. The well-formedness condition is preserved by step 4 because this step does not change $\bigcup \mathsf{Comp}(S)$ or $\bigcup \mathsf{Split}(S)$. Step 5 does not violate this condition either because it sets the components of $\emptyset_d$ to $\emptyset$.

*ii)* To see that the resulting diagram is simple, we show that it satisfies conditions *a), ..., f)* of Definition 6. After step 2 of the algorithm, the resulting factor-diagram has no cycles of length 2 or more, there are only potentially some self-cycles. These are eliminated in step 3 and no further edges are introduced. Hence, *a)* holds. For each of the following condition, they are enforced in certain step and not violated afterwards, according to the following table:

| b) | c) | d) | e) | f) |
|---|---|---|---|---|
| 3. | 4. | 4. | 4. | 5. |

*iii)* We show that semantics is preserved when executing each sequence of steps $1, \ldots, k$ for $2 \leq k \leq 5$, that is, each step preserves the semantics provided that it is executed after the previous steps.

$k = 2$. Each equality introduced into $\rho$ is a semantic consequence of the diagram, because

1.1 $\overline{\alpha}(S_1) \subseteq S_2$ and $\overline{\alpha}(S_2) \subseteq \overline{\alpha}(S_1)$,

1.2 $\overline{\alpha}(S_1) \subseteq \overline{\alpha}(\emptyset_d) = \emptyset$,

1.3 $\overline{\alpha}(S_1) \subseteq \bigcup \emptyset = \emptyset$,

1.4 $\overline{\alpha}(S_1) \cap \overline{\alpha}(S_2) = \emptyset$ for $\overline{\alpha}(S_1) = \overline{\alpha}(S_2)$ so $\overline{\alpha}(S_2) = \emptyset$, or

1.5 $\overline{\alpha}(S_1) \cap \overline{\alpha}(S_2) = \emptyset$, for $\overline{\alpha}(S_1) = \overline{\alpha}(S_0)$, and $\overline{\alpha}(S_2) \subseteq \overline{\alpha}(S_0)$, so again $\overline{\alpha}(S_2) = \emptyset$,

1.6 $\overline{\alpha}(S_1) \subseteq S$ and $\overline{\alpha}(S) \subseteq \bigcup \{\overline{\alpha}(S_1)\} = \overline{\alpha}(S_1)$.

It follows that the condition on equality of sets, as well as the conditions on $\mathsf{CInf}$, $\mathsf{CSup}$, $\mathsf{Sons}$, $\Phi$, $\mathsf{Comp}$ are all semantically equivalent when applied to the original and the factor diagram. The only semantic condition which can be lost in factor-diagram construction is $\mathsf{disj}_{\mathcal{D},S_0}(S_1, S_2)$ when $S_1$ and $S_2$ nodes are merged, that is, when $(S_1, S_2) \in \rho$. However, in this case the disjointness condition follows from $\overline{\alpha}(S_1) = \emptyset$, which is enforced in 1.3. Therefore, for the particular relation constructed in step 1, factor-diagram is an equivalence preserving transformation.

$k = 3$. We need to show that no information is lost by removing $\emptyset_d$ and $S$ from the sons, as well as split and complete views of $S$. Clearly, removing $S$ and $\emptyset_d$ from $\mathsf{Sons}(S)$ does not change the subset conditions because $\emptyset \subseteq \overline{\alpha}(S)$ and $\overline{\alpha}(S) \subseteq \overline{\alpha}(S)$. Eliminating $\emptyset_d$ from $Q \in \mathsf{Comp}(S)$ is justified because the view has the same semantics with or without $\emptyset_d$. Dropping a view $Q \in \mathsf{Comp}(S)$ for $S \in Q$ is justified because in that case $\overline{\alpha}(S) \subseteq \bigcup_{S_1 \in Q} \overline{\alpha}(S_1)$ holds trivially. Eliminating $\emptyset_d$ from $Q \in \mathsf{Split}(S)$ is justified because intersection with empty set is always empty, so this condition does not bring any new information. Finally, dropping a $Q \in \mathsf{Split}(S)$ with $S \in Q$ is justified because condition (C) implies that in such case $Q \subseteq \{\emptyset_d, S\}$ so the Split condition is trivial.

$k = 4$. Removing $\{\emptyset_d\}$ from $\mathsf{Split}(S)$ preserves semantics because such view carries no information. Similarly, because all maximal views are preserved, removing their subsets does not change the semantics. For $\mathsf{Comp}(S)$, we consider two cases. In first case $\emptyset \notin \mathsf{Comp}(S)$. In this case, removing $\emptyset$ does not have any effect, and it is sound to remove all non-minimal views because they are implied by the minimal views. The second case is $\emptyset \in \mathsf{Comp}(S)$. By condition D) on the step 1, we know that $Q \neq \emptyset$ after the step 2, and the only node removed in step 3 is $\emptyset_d$, so it must have been the case that $Q = \{\emptyset_d\}$ after step 2. By condition D), we then have $S = \emptyset_d$. Because the semantic condition on $\mathsf{Comp}$ for $Q = \emptyset$ reduces to $\overline{\alpha}(S) = \emptyset$, this condition brings no new information, so we can remove it.

$k = 5$. Because $\overline{\alpha}(\emptyset_d) =$, $\mathsf{CSup}(\emptyset_d) = 0$ does not change semantics, similarly for $\Phi(\emptyset_d) = \emptyset$. We also know that $\mathsf{Sons}(S) \subseteq \{\emptyset_d\}$ because this condition is ensured by step 2 and is not violated afterwards. Because we have already observed that the diagram is well-formed, we conclude $\mathsf{Comp}(S) \subseteq \{\emptyset_d\}$ and $\mathsf{Split}(S) \subseteq \{\emptyset_d\}$, so setting these values to $\emptyset$ does not change the semantics. ∎

**Theorem 2** *Omitting any one out of three conditions from Definition 9 (1. being tree-shaped, 2. having independent views, and 3. having independent signatures) yields a class of diagrams whose satisfiability is NP-hard.*

*Proof.* Suppose that at least one of the three conditions does not apply to a class of i-diagrams. We then give a reduction from the problem 3COL to the satisfiability of i-diagrams in this class. Here 3COL denotes the NP-complete problem of deciding, given an indirected graph, whether the graph can be colored using 3 colors such that adjacent nodes have different colors [41, Page 275].

Given a graph $(N, E)$ where $E \subseteq N \times N$ is a symmetric irreflexive relation, we first build the i-diagram $\mathcal{D}$ defined as follows: $\mathbb{S} = N \cup \{U\}$, $\mathsf{CInf}(U) = \mathsf{CSup}(U) = 3$, $\mathsf{Sons}(U) = N$, $\mathsf{Split}(U) = \{\{n\} | n \in N\}$, and $\mathsf{Comp}(U) = \Phi(U) = \emptyset$. For all $n \in N$ we let $\mathsf{CInf}(n) = \mathsf{CSup}(n) = 1$ and let $\mathsf{Sons}(n) = \mathsf{Split}(n) = \mathsf{Comp}(n) = \Phi(n) = \emptyset$.

Each model of this diagram is a triple $(\Delta, \alpha, \Xi)$ such that $|\overline{\alpha}(U)| = 3$ and for all $n$, $\overline{\alpha}(n)$ is a singleton included in $\overline{\alpha}(U)$. If we consider $\overline{\alpha}(U)$ as a set of three colors, then in each model with this property, $\overline{\alpha}(n)$ indicates the color of node $n$.

Then for each edge $(n_1, n_2) \in E$ we encode the fact that $n_1$ and $n_2$ must have different colors by enforcing the property $\overline{\alpha}(n_1) \cap \overline{\alpha}(n_2) = \emptyset$ on the models of $\mathcal{D}$. We encode this constraint in different ways depending on the class of i-diagrams:

- If $\mathcal{D}$ allows dependent signatures, we introduce a fresh predicate symbol $P_{n_1, n_2}$, and add $(+P_{n_1, n_2})$ to $\Phi(n_1)$, and $(-P_{n_1, n_2})$ to $\Phi(n_2)$.
- If $\mathcal{D}$ allows dependent views, we add $\{n_1, n_2\}$ to $\mathsf{Split}(U)$.
- If $\mathcal{D}$ allows multiple fathers, then we simulate dependent views by introducing a new node $m(n_1, n_2)$. We let $\mathsf{CInf}(m(n_1, n_2)) = \mathsf{CSup}(m(n_1, n_2)) = 2$, $\mathsf{Sons}(m(n_1, n_2)) = \{n_1, n_2\}$, $\mathsf{Split}(m(n_1, n_2)) = \mathsf{Comp}(m(n_1, n_2)) = \{\{n_1, n_2\}\}$, and $\Phi(m(n_1, n_2)) = \emptyset$. We then remove $n_1, n_2$ from $\mathsf{Sons}(U)$, and add $m(n_1, n_2)$ to $\mathsf{Sons}(U)$ instead.

None of these constructions violates more than one of the three considered restrictions. It is straightforward to verify that the diagram is satisfiable iff the graph is colorable, and that the construction of $\mathcal{D}$ can be done in polynomial time. This proves that the satisfiability of i-diagrams with any of the three restrictions removed is NP-hard. ■

### A.2 Termination of System $\mathcal{R}$

LEMMA 17 (Invariants of $\mathcal{R}$). *For every $k \in [1..12]$ the rule $R_k$ preserves $\bigwedge_{j=1..(k-1)} \mathcal{C}_j$ or returns $\bot_d$.*

**Proof.** We analyze each rule $R_k$, for two i-trees $\mathcal{T}$ and $\mathcal{T}'$ such that $\mathcal{T} \xrightarrow{R_k} \mathcal{T}'$, assuming that $\mathcal{T}$ satisfies $\bigwedge_{j=1..(k-1)} \mathcal{C}_j$, and, more precisely, $\mathcal{T} \xrightarrow{\text{spot}}_{R_k} \mathcal{T}'$, where *spot* are variable names as they appear in the definition of $\mathcal{R}$.

1. (**DnPhi**). Trivial.
2. (**Unsat**). $(\mathcal{C}_1)$ does not depend on $\mathsf{CSup}$.
3. (**UpSup**). If $n = 0$, $(\mathcal{C}_2)$ is trivialy true for $S$ in $\mathcal{D}'$. If $n > 0$, by $(c_3)$ $\mathsf{CSup}(S) > 0$ and we have already $\forall P \in \mathsf{PN}.\{+P, -P\} \not\subseteq \Phi(S)$ and since $\Phi' = \Phi$, $(\mathcal{C}_2)$ holds in $\mathcal{T}'$.
4. (**UpInf**). Neither $(\mathcal{C}_1),(\mathcal{C}_2)$ nor $(\mathcal{C}_3)$ depend on $\mathsf{CInf}$.
5. (**Error**). $\mathcal{T}' = \bot_d$
6. (**DnInf**). Only $(\mathcal{C}_4)$ and $(\mathcal{C}_5)$ depends on $\mathsf{CInf}$.
   - $(\mathcal{C}_5)$ is maintained for $S$ in $\mathcal{T}'$ because, noticing that $(\mathcal{C}_5)$ is maintained for $S' \neq S$, we have
   $$\begin{aligned}\mathsf{CInf}'(S) &= \mathsf{CInf}(S') - \Sigma(\mathsf{CSup}(Q_0)) \\ &\leq \mathsf{CSup}(S') - \Sigma(\mathsf{CSup}(Q_0)) \quad \text{(by } \mathcal{C}_5\text{)} \\ &\leq \mathsf{CSup}(S) \quad \text{(by } \mathcal{C}_3\text{)}\end{aligned}$$
   - To prove that $(\mathcal{C}_4)$ is maintained in $\mathcal{T}'$ we need to check that $(\mathcal{C}_4)$ is maintained for $S$, which is trivial by $(c_6)$, and that $(\mathcal{C}_4)$ is maintained for the father $S'$ of $S$ and the views $Q \in \mathsf{Split}(S')$ containing $S$. By property of independent views there exists only one such view $Q = (\{S\} \uplus Q'_0)$ such that $Q'_0 \subseteq Q_0$ and
   $$\begin{aligned}\Sigma\mathsf{CInf}'(Q) &= \mathsf{CInf}(S) + \Sigma\mathsf{CInf}(Q'_0) \\ &= (\mathsf{CInf}(S') - \Sigma\mathsf{CSup}(Q_0)) + \Sigma\mathsf{CInf}(S'_0) \\ &= \mathsf{CInf}(S') - \Sigma\mathsf{CSup}(Q_0 - Q'_0) \\ &\quad + \Sigma_{S'_0 \in Q'_0}(\mathsf{CInf}(S'_0) - \mathsf{CSup}(S'_0)) \\ &\leq \mathsf{CInf}(S') - \Sigma\mathsf{CSup}(Q_0 - Q'_0) \quad \text{(by } \mathcal{C}_5\text{)} \\ &\leq \mathsf{CInf}(S') \\ &= \mathsf{CInf}'(S')\end{aligned}$$

7. (**DnSup**). Only $(\mathcal{C}_2)$, $(\mathcal{C}_3)$ and $(\mathcal{C}_5)$ depend on $\mathsf{CSup}$. $(\mathcal{C}_2)$ is maintained thanks to $(c_7)$ as for the case of rule **UpSup**.
   - $(\mathcal{C}_5)$ is maintained for $S$ in $\mathcal{T}'$ because, noticing that $(\mathcal{C}_5)$ is maintained for $S' \neq S$ we have
   $$\begin{aligned}\mathsf{CSup}'(S) &= \mathsf{CSup}(S') - \Sigma(\mathsf{CInf}(Q_0)) \\ &\geq \mathsf{CInf}(S') - \Sigma(\mathsf{CInf}(Q_0)) \quad \text{(by } \mathcal{C}_5\text{)} \\ &\geq \mathsf{CInf}(S) \quad \text{(by } \mathcal{C}_4\text{)}\end{aligned}$$
   - To prove that $(\mathcal{C}_3)$ is maintained in $\mathcal{T}'$ we need to check that $(\mathcal{C}_3)$ is maintained for $S$, which is trivial by $(c_7)$, and that $(\mathcal{C}_3)$ is maintained for the father $S'$ of $S$ and the views $Q \in \mathsf{Comp}(S')$ containing $S$. By property of independent views there exists only one such view $Q = (\{S\} \uplus Q_0)$ and
   $$\begin{aligned}\Sigma\mathsf{CSup}'(Q) &= \mathsf{CSup}'(S) + \Sigma\mathsf{CSup}(Q_0) \\ &= (\mathsf{CSup}(S') - \Sigma\mathsf{CInf}(Q_0)) + \Sigma\mathsf{CSup}(S_0) \\ &= \mathsf{CInf}(S') + \Sigma_{S_0 \in Q_0}(\mathsf{CSup}(S_0) - \mathsf{CInf}(S_0)) \\ &\geq \mathsf{CInf}(S') \quad \text{(by } \mathcal{C}_5\text{)} \\ &= \mathsf{CInf}'(S')\end{aligned}$$

8. (**CCmp**). Only $(\mathcal{C}_3)$ and $(\mathcal{C}_6)$ depend on $\mathsf{Comp}$.
   - $(\mathcal{C}_3)$ is maintained for $S, Q$ because
   $$\begin{aligned}\mathsf{CSup}'(S) &= \mathsf{CSup}(S) \\ &\leq \Sigma(\mathsf{CInf}(Q)) \quad \text{(by } a_8\text{)} \\ &\leq \Sigma(\mathsf{CSup}(Q)) \quad \text{(by } \mathcal{C}_5\text{)} \\ &= \Sigma(\mathsf{CSup}'(Q))\end{aligned}$$
   - $(\mathcal{C}_6)$ is maintained for $S, Q$ and all $S_0 \in Q$ because
   $$\begin{aligned}\mathsf{CInf}'(S) &= \mathsf{CInf}(S) \\ &\leq \mathsf{CSup}(S) \quad \text{(by } \mathcal{C}_5\text{)} \\ &\leq \Sigma(\mathsf{CInf}(Q)) \quad \text{(by } a_8\text{)} \\ &= \mathsf{CInf}(S_0) + \Sigma(\mathsf{CInf}(Q - \{S_0\})) \\ &\leq \mathsf{CInf}(S_0) + \Sigma(\mathsf{CSup}(Q - \{S_0\})) \quad \text{(by } \mathcal{C}_5\text{)} \\ &= \mathsf{CInf}'(S_0) + \Sigma(\mathsf{CSup}'(Q - \{S_0\}))\end{aligned}$$

(*Remark*). If ever we use a simplification afterwards, as indicated by the star in figure 10, it can only consists in removing a complete view $Q'$ such that $Q' \subsetneq Q$. This operation trivially maintains $\bigwedge_{j=1..(7)} \mathcal{C}_j$ because in every properties of consistency where complete views appear they are universally quantified.

9. (**CSplit**). Only $(\mathcal{C}_4)$ and $(\mathcal{C}_7)$ depend on $\mathsf{Split}$.
   - $(\mathcal{C}_4)$ is maintained for $S, Q$ because
   $$\begin{aligned}\mathsf{CInf}'(S) &= \mathsf{CInf}(S) \\ &\geq \Sigma(\mathsf{CSup}(Q)) \quad \text{(by } a_9\text{)} \\ &\geq \Sigma(\mathsf{CInf}(Q)) \quad \text{(by } \mathcal{C}_5\text{)} \\ &= \Sigma(\mathsf{CInf}'(Q))\end{aligned}$$
   - $(\mathcal{C}_7)$ is maintained for $S, Q$ and all $S_0 \in Q$ because
   $$\begin{aligned}\mathsf{CSup}'(S) &= \mathsf{CSup}(S) \\ &\geq \mathsf{CInf}(S) \quad \text{(by } \mathcal{C}_5\text{)} \\ &\geq \Sigma(\mathsf{CSup}(Q)) \quad \text{(by } a_9\text{)} \\ &= \mathsf{CSup}(S_0) + \Sigma(\mathsf{CSup}(Q - \{S_0\})) \\ &\geq \mathsf{CSup}(S_0) + \Sigma(\mathsf{CInf}(Q - \{S_0\})) \quad \text{(by } \mathcal{C}_5\text{)} \\ &= \mathsf{CSup}'(S_0) + \Sigma(\mathsf{CInf}'(Q - \{S_0\}))\end{aligned}$$

(*Remark*). If ever we use a simplification afterwards, as indicated by the star in figure 10, it can only consists in removing a split view $Q'$ such that $Q' \subsetneq Q$. This operation trivially maintains $\bigwedge_{j=1..8} \mathcal{C}_j$ because in every properties of consistency where split views appear they are universally quantified.

10. (**UpPhi**). Only ($\mathcal{C}_1$) and ($\mathcal{C}_2$) depend on $\Phi$

    - ($\mathcal{C}_1$) is maintained for $S$ and all $S' \in Q$ because

      $$\begin{aligned}\Phi'(S) &= \Phi(S) \cup \bigcap \Phi(Q) & \text{(by } b_{10}) \\ &= \Phi(S) \cup (\Phi(S') \bigcap \Phi(Q-\{S'\})) \\ &\subseteq \Phi(S) \cup \Phi(S') \\ &\subseteq \Phi(S') & \text{(by } \mathcal{C}_1) \\ &= \Phi'(S')\end{aligned}$$

    - We can also prove that ($\mathcal{C}_2$) is maintained for $S$. Suppose there exists $P$ in PN such that $\{+P, -P\} \in \Phi'(S)$. Then we distinguish three cases.

      - if $\{+P, -P\} \subseteq \Phi(S)$, by $\mathcal{C}_2$, $\mathsf{CSup}(S) = 0$
      - if $\{+P, -P\} \cap \Phi(S) = \emptyset$, then $\{+P, -P\} \in \bigcap \Phi(Q)$. Then for all $S' \in Q$, $\mathsf{CSup}(S') = 0$ by $\mathcal{C}_2$. Then by $\mathcal{C}_3$, $\mathsf{CSup}(S) = 0$
      - if $\{+P, -P\} \cap \Phi(S) = \pm P$ for one atom $\pm P \in \{+P, -P\}$. Then the opposite atom $\mp P$ belong $\bigcap \Phi(Q)$ and by $\mathcal{C}_1$, $\pm P$ belong to $\bigcap \Phi(Q)$. By $\mathcal{C}_2$, each node $S' \in Q$ is such that $\mathsf{CSup}(S') = 0$ and by $\mathcal{C}_3$, $\mathsf{CSup}(S) = 0$

      In the three cases $\mathsf{CSup}'(S) = \mathsf{CSup}(S) = 0$.

11. (**Void**). If $S$ is the root, the resulting i-tree $\mathcal{T}'$ is such that $\mathbb{S}' = \{\mathsf{SN}\} = \{\emptyset_d\}$ and $\mathcal{C}_i$ is trivial for all $i \in [1..11]$. Otherwise, we have to check that removing the node $S$ from the sons of the father of $S$ (as indicated in the step 3 of the procedure **simplify**) maintains $\bigwedge_{j=1..10} \mathcal{C}_j$. We denote by $S_0$ the father of $S$ and by $Q_0$ the split view of $S_0$ containing $S$. If $S$ is also contained in a complete view of $S_0$ we denote by $C_0$ this complete view.

    - ($\mathcal{C}_1$) and ($\mathcal{C}_2$) are trivially maintained.
    - ($\mathcal{C}_3$) is maintained because, if $C_0$ exists and $C'_0 = C_0 - \{S\}$ is not empty

      $$\begin{aligned}\mathsf{CSup}'(S_0) &= \mathsf{CSup}(S_0) \\ &\leq \Sigma \mathsf{CSup}(C_0) & (\text{by } \mathcal{C}_3) \\ &= \Sigma \mathsf{CSup}(C'_0) & (\text{by } a_{12}) \\ &= \Sigma \mathsf{CSup}'(C'_0)\end{aligned}$$

    - ($\mathcal{C}_4$) is maintained because, if $Q'_0 = Q_0 - \{S\}$ is not empty

      $$\begin{aligned}\mathsf{CInf}'(S_0) &= \mathsf{CInf}(S_0) \\ &\geq \Sigma \mathsf{CInf}(Q_0) & (\text{by } \mathcal{C}_4) \\ &\geq \Sigma \mathsf{CInf}(Q'_0) \\ &= \Sigma \mathsf{CInf}'(Q'_0)\end{aligned}$$

    - ($\mathcal{C}_5$) is maintained for the node $\emptyset'_d = S \cup \emptyset_d$ of $\mathcal{T}'$ because by ($a_{12}$) and ($\mathcal{C}_5$), $\mathsf{CInf}(S) \leq \mathsf{CSup}(S) = 0$, by simplicity and ($\mathcal{C}_5$), $\mathsf{CInf}(\emptyset_d) \leq \mathsf{CSup}(\emptyset_d) = 0$, and therefore $\mathsf{CInf}'(\emptyset'_d) = Max(\mathsf{CInf}(S), \mathsf{CInf}(\emptyset_d)) = 0 \leq \mathsf{CSup}'(\emptyset'_d)$.

    - ($\mathcal{C}_6$) is maintained for $S_0, C_0$ and every $S_1 \in C_0$ such that $S_1 \neq S$ because

      $$\begin{aligned}\mathsf{CInf}'(S_1) &= \mathsf{CInf}(S_1) \\ &\geq \mathsf{CInf}(S_0) - \Sigma(\mathsf{CSup}(C_0-\{S_1\})) & \text{(by } \mathcal{C}_6) \\ &= \mathsf{CInf}(S_0) - \Sigma(\mathsf{CSup}(C_0-\{S\}-\{S_1\})) & \text{(by } a_{12}) \\ &= \mathsf{CInf}'(S_0) - \Sigma(\mathsf{CSup}'(C'_0-\{S_1\}))\end{aligned}$$

    - ($\mathcal{C}_7$) is maintained for $S_0, Q_0$ and every $S_1 \in Q_0$ such that $S_1 \neq S$ because

      $$\begin{aligned}\mathsf{CSup}'(S_1) &= \mathsf{CSup}(S_1) \\ &\leq \mathsf{CSup}(S_0) - \Sigma(\mathsf{CInf}(C_0-\{S_1\})) & \text{(by } \mathcal{C}_7) \\ &\leq \mathsf{CSup}(S_0) - \Sigma(\mathsf{CInf}(C_0-\{S\}-\{S_1\})) \\ &= \mathsf{CSup}'(S_0) - \Sigma(\mathsf{CInf}'(C'_0-\{S_1\}))\end{aligned}$$

    - ($\mathcal{C}_8$) is maintained for $Q_0$ because

      $$\begin{aligned}\mathsf{CSup}'(S_0) &= \mathsf{CSup}(S_0) \\ &> \Sigma(\mathsf{CInf}(Q_0)) & (\text{ by } \mathcal{C}_8) \\ &\geq \Sigma(\mathsf{CInf}(Q_0-\{S_0\})) \\ &= \Sigma(\mathsf{CInf}'(Q'_0))\end{aligned}$$

    - ($\mathcal{C}_9$) is maintained for $C_0$ (if exists and $C'_0 = C_0 - \{S_0\} \neq \emptyset$) because

      $$\begin{aligned}\mathsf{CInf}'(S_0) &= \mathsf{CInf}(S_0) \\ &< \Sigma(\mathsf{CSup}(Q_0)) & (\text{ by } \mathcal{C}_9) \\ &= \Sigma(\mathsf{CSup}(Q_0-\{S_0\})) & (\text{ by } a_{11}) \\ &= \Sigma(\mathsf{CSup}'(Q'_0))\end{aligned}$$

    - ($\mathcal{C}_{10}$) is maintained for $C_0$ (if exists and $C'_0 = C_0-\{S_0\} \neq \emptyset$) because

      $$\begin{aligned}\bigcap \Phi'(C'_0) &= \bigcap \Phi(C_0-\{S_0\}) \\ &\subseteq \bigcap \Phi(C_0) \\ &\subseteq \Phi(S_0) & (\text{ by } \mathcal{C}_{10}) \\ &= \Phi'(S_0)\end{aligned}$$

12. (**Equal**). Before to merge $S$ and $S'$ for $\{S'\} \in \mathsf{Comp}(S)$ we have

    - $\mathsf{CInf}(S) = \mathsf{CInf}(S')$ by $\mathcal{C}_4$ and $\mathcal{C}_6$
    - $\mathsf{CSup}(S) = \mathsf{CSup}(S')$ by $\mathcal{C}_3$ and $\mathcal{C}_7$
    - $\Phi(S) = \Phi(S')$ by $\mathcal{C}_1$ and $\mathcal{C}_{10}$

    Therefore all the properties $\mathcal{C}_i$ for $i = 1..10$ are trivially maintained.

∎

### A.3 Model Construction

**Lemma 6 (Model Construction)** *If an i-tree $\mathcal{T}$ is weakly consistent, then we can construct a model for $\mathcal{T}$.*

*Proof.* Let $\mathcal{T}$ be a weakly consistent i-tree.

We construct a model $(\Delta, \alpha, \Xi)$ by first constructing a partial model $(\Delta, \alpha)$ for all parts of $\mathcal{T}$ except $\Phi$, and then extending $(\Delta, \alpha)$ with $\Xi$ to satisfy $\Phi$.

**Constructing $(\Delta, \alpha)$.** We write $(\Delta, \alpha) \models \mathcal{T}$ to denote that $\Delta$ and $\alpha$ satisfy those conditions on $\mathbb{S}$, Sons, Split, Comp, CSup, CInf that do not mention $\Phi$ in Definition 2. To show we can construct $(\Delta, \alpha)$ such that $(\Delta, \alpha) \models \mathcal{T}$ we prove by induction on $n$ the following more general claim.

CLAIM 1. *For every i-tree $\mathcal{T}$ of height $n$ with root $S_R$:*

$$\forall k \in [\mathsf{CInf}(S_R), \mathsf{CSup}(S_R)].$$
$$\exists (\Delta, \alpha). \ (\Delta, \alpha) \models \mathcal{D} \wedge |\overline{\alpha}(S_R)| = |\Delta| = k$$

If $n = 1$, the claim holds by $\mathcal{C}_5$ taking

$$\Delta = \overline{\alpha}(S_R) = \{1, \ldots, k\}$$

For $n > 1$, consider an i-tree $\mathcal{T}$ with root $S_R$ and $k \in [\mathsf{CInf}(S_R), \mathsf{CSup}(S_R)]$. By examining the constraints in $\mathcal{T}$, we choose the cardinalities for subtrees of $\mathcal{T}$, use the induction hypothesis to construct models for subtrees, and paste the models for subtrees into a model for $\mathcal{T}$. We decompose this process into three steps:

1. For each $C \in \mathsf{Comp}(S_R)$, consider a subtree $\mathcal{T}_C$ built from $\mathcal{T}$ by removing all sons outside $\bigcup C$. We construct a model $\mathcal{M}_C$ for $\mathcal{T}_C$ of cardinality $k$.
2. For each remaining $Q \in \mathsf{Split}(S_R)$ where $Q \not\subseteq \bigcup \mathsf{Comp}(S_R)$, consider a subtree $\mathcal{T}_Q$ with the same root $S_R$ but without the sons outside $\bigcup Q$. We construct a model $\mathcal{M}_Q$ for $\mathcal{T}_C$ of cardinality $k$.
3. Because the constructed models have the same cardinality, we can easily merge them to obtain a model for $\mathcal{T}$.

**Step 1.** Let $C \in \mathsf{Comp}(S_R)$, and $\mathbb{C} \subseteq \mathsf{Split}(S_R)$ such that $C = \cup \mathbb{C}$. For each $Q \in \mathbb{C}$, let $t_Q \stackrel{def}{=} \min(k, \Sigma(\mathsf{CSup}[Q]))$. Using $k \geq \mathsf{CInf}(S_R)$ and $\mathcal{C}_4$ we can show

$$\Sigma(\mathsf{CInf}[Q]) \leq t_Q \leq \Sigma(\mathsf{CSup}[Q]) \qquad (H1)$$

To each node $S \in Q$ we can therefore assign an integer $K(S) \in [\mathsf{CInf}(S), \mathsf{CSup}(S)]$ such that $t_Q = \Sigma(K[Q])$. Let $\mathcal{T}_S$ be the sub-i-tree of $\mathcal{T}$ rooted at $S$. By induction hypothesis, let $\mathcal{M}_S$ be the model of $\mathcal{T}_S$ of cardinality $K(S)$. We can then take the disjoint union of these models to construct a model $\mathcal{M}_{M_Q}$ of size $t_Q$ for the forest $\bigcup_{S \in Q} \mathcal{T}_S$.

For all $Q \in \mathbb{C}$, we have $t_Q \leq k$ by definition of $t_Q$ and $k$. We can also prove that $k \leq \Sigma_{Q \in \mathbb{C}} t_Q$. Indeed, if there exists a $Q_0 \in \mathbb{C}$ such that $t_{Q_0} = k$, this is trivial. Otherwise, because $\mathcal{T}$ is weakly consistent, from $\mathcal{C}_3$ we can show that $k \leq \mathsf{CSup}(S) \leq \Sigma(\mathsf{CSup}[C])$ because

$$\Sigma(\mathsf{CSup}[C]) = \sum_{Q \in \mathbb{C}}(\Sigma(\mathsf{CSup}[Q])) = \sum_{Q \in \mathbb{C}}(t_Q).$$

We finally obtain

$$\max_{Q \in \mathbb{C}} t_Q \leq k \leq \sum_{Q \in \mathbb{C}} t_Q \qquad (H2)$$

Thanks to $(H2)$, we can build a model for the i-tree $\mathcal{T}_C$, as follows. We start with the disjoint union of models $\mathcal{M}_Q$ for $\mathcal{T}_Q$ for $Q \in \mathbb{C}$. This model has cardinality $\Sigma_{Q \in \mathbb{C}} t_Q$. Then, we rename elements from different models to be identical to elements from other models. Such merging is possible as long as there is no model whose domain contains the domains of all others, so we can reach any cardinality $k$ for $\max_{Q \in \mathbb{C}} \leq k$.

REMARK 2. (Freedom in the choice of $\{t_i\}_{i \in [1..n]}$) In Section 7 we enforce some additional properties on models using a different choice of $t_Q$. Such construction is possible whenever $t_i$ satisfy $(H1)$ and $(H2)$. Moreover, if $K(S)$ denotes some chosen cardinality for each node $S$, and the values $K(S)$ satisfy certain assumptions, then we can enforce additional properties when merging the models $\mathcal{M}_Q$ corresponding for $Q \in \mathsf{Split}(S)$. The following two cases are of interest.

1. If $\sum_{Q \in \mathbb{C}} t_Q > K(S)$, we can chose any pair of different split views $Q_1, Q_2 \in \mathbb{C}$, and two elements $x_1$ from $\mathcal{M}_{Q_1}$ and $x_2$ from $\mathcal{M}_{Q_2}$ and decide to merge them.
2. If $\mathbb{C} = \{Q_0\} \uplus \mathbb{C}_0$ and $\max_{Q \in \mathbb{C}_0} t_Q < K(S)$, we can chose any element in the model $\mathcal{M}_{Q_0}$ and decide not to merge it with any of the elements of the models $\mathcal{M}_{Q'}$ for $Q' \in \mathbb{C}_0$.

**Step 2.** Let $Q \in \mathsf{Split}(S_R)$, such that $Q$ is not included in any complete view. We construct a model $\mathcal{M}_Q$ of size $k$ for the i-forest $\mathcal{T}_T$ by first building a model of size $K(S') = \mathsf{CInf}(S')$ for each $S' \in Q$. Because $k \geq \mathsf{CInf}(S) \geq \Sigma(\mathsf{CInf}[Q])$, by $\mathcal{C}_4$, the disjoint union of these models has cardinality smaller than $k$. By adding the correct number of fresh elements to $\Delta$, we obtain a model of cardinality $k$.

REMARK 3. (Existence of fresh elements) In Section 7 we use the following property: For all $Q \in \mathsf{Split}(S)$ and $Q \notin \cup \mathsf{Comp}(S)$, if $\Sigma(\mathsf{CInf}[Q]) < K(S)$, then there exists a model such that $\overline{\alpha}(S)$ contains an element which does not belong to any $\overline{\alpha}(S')$ for any of the sons of $S$.

**Step 3.** We can apply an arbitrary bijection $\sigma_C : \Delta_C \to [1..n]$ to each model $\mathcal{M}_C$ constructed as previously described before to build a model for the entire i-tree. We let $\overline{\alpha}(S_R) = [1..n]$ and for all $S \neq S_R$, $\overline{\alpha}(S) = \overline{\alpha}_C(S)$ where $C$ is the view containing an ancestor of $S$ in $\mathsf{Split}(S_R)$ or $\mathsf{Comp}(S_R)$.

REMARK 4. (Freedom in the choice of $\sigma$.) If we know that $K(S) > 0$, for any pair $S_1, S_2$ of sons of $S$ such that $S_1, S_2$ belong neither to the same split view nor to the same complete view, for each choice of elements $x_1, x_2$ in the models $\mathcal{T}_{S_1}$ and $\mathcal{T}_{S_2}$ we can choose $\sigma_1, \sigma_2$ such that $\sigma_1(x_1) = \sigma_2(x_2) \stackrel{def}{=} x$ and the resulting i-tree will be such that $x \in \overline{\alpha}(S_1) \cap \overline{\alpha}(S_2)$.

**Extending the model with $\Xi$.** Let $(\Delta, \alpha) \models \mathcal{T}$. Then for each $P \in \mathsf{PN}$, define

$$\Xi(P) = \bigcup \{\overline{\alpha}(S) \mid S \in \mathbb{S} \wedge P \in \Phi(S)\}$$

Then $(+P) \in \Phi(S) \Rightarrow \overline{\alpha}(S) \subseteq \Xi(P)$ holds by construction, it remains to show $(-P) \in \Phi(S) \Rightarrow \overline{\alpha}(S) \subseteq \Xi(P)^c$ for every node $S$. Consider $S_1 \in \mathbb{S}$ such that $(-P) \in \Phi(S_1)$. If $S_1 = \emptyset_d$, then $\overline{\alpha}(S_1) = \emptyset$, so the condition trivially holds. Similarly, if $(+P) \in \Phi(S_1)$, then by $\mathcal{C}_2$, $\mathsf{CSup}(S_1) = 0$ so $\overline{\alpha}(S_1) = \emptyset$ and the condition holds. Otherwise, assume $(+P) \notin \Phi(S_1)$. For the sake of contradiction suppose that there exists an element $x \in \overline{\alpha}(S_1)$, $x \in \Xi(P)$. By definition of $\Xi(P)$, there exists a node $S_2 \neq S_1$ such that $x \in \overline{\alpha}(S_2)$ and $(+P) \in \Phi(S_2)$. By the condition on independent signatures, one of the followig two cases applies.

1. $\mathsf{disj}^*_\mathcal{T}(S_1, S_2)$. Then $\overline{\alpha}(S_1) \cap \overline{\alpha}(S_2) = \emptyset$ by the semantics of i-diagrams, which is a contradiction with $x \in \overline{\alpha}(S_1)$ and $x \in \overline{\alpha}(S_2)$.
2. $S_1$ and $S_2$ have compatible signatures. Then there exists a node $S$ such that $S_1 \stackrel{*}{\leadsto} S$, $S_2 \stackrel{*}{\leadsto} S$ and $\mathsf{Sig}(S_1) \cap \mathsf{Sig}(S_2) \subseteq \mathsf{Sig}(S)$. Because $(-P) \in \Phi(S_1)$, $(+P) \in \mathsf{Sig}(S_1)$, and because $(-P) \in \Phi(S_2), P \in \mathsf{Sig}(S_2)$. Therefore $P \in \mathsf{Sig}(S)$. We have two cases:
   (a) $(+P) \in \Phi(S)$. By $\mathcal{C}_1$, then $(+P) \in \Phi(S_1)$, a contradiction.
   (b) $(-P) \in \Phi(S)$. By $\mathcal{C}_2$, then $(-P) \in \Phi(S_2)$. By $\mathcal{C}_2$ then $\mathsf{CSup}(S_2) = 0$, so $\overline{\alpha}(S_2) = \emptyset$, a contradiction with $x \in \overline{\alpha}(S_2)$.

We have reached the contradiction in each case, so we conclude $\overline{\alpha}(S_1) \subseteq \Xi(P)^c$. ∎

### A.4 Details of the Proofs for Subsumption Completeness

#### A.4.1 Refinemenents of Lemma 6

According to the remarks in the proof of Lemma 6, if an i-tree $\mathcal{T}$ is weakly consistent, there exists a choice of cardinalities $K : \mathbb{S} \to \mathbb{N}$, such that we can build a model $(\Delta, \alpha, \Xi)$ for $\mathcal{T}$ with the property $|\overline{\alpha}(S)| = K(S)$ for all $S \in \mathbb{S}$. For a fixed choice of cardinalities $K$, we can, in certain cases, enforce some additional properties by choosing which element we merge in the steps 1 and 3 of the construction. The three following Lemmas are based on this idea.

LEMMA 18 (Non-empty intersection (1)). *Let $S_1$ and $S_2$ be nodes in a weakly consistent i-tree $\mathcal{T}$ is such that*

$$\mathsf{CInf}(S_1) > 0 \wedge S_1 \stackrel{*}{\leadsto} S'_1 \wedge S'_1 \in Q_1 \wedge$$
$$\mathsf{CInf}(S_2) > 0 \wedge S_2 \stackrel{*}{\leadsto} S'_2 \wedge S'_2 \in Q_2$$

*for some $S, S'_1, S'_2 \in \mathbb{S}$, $Q_1, Q_2 \in \mathsf{Split}(S)$ where $Q_1 \neq Q_2$ and $\neg(\exists C \in \mathsf{Comp}(S). Q_1 \subseteq C \wedge Q_2 \subseteq C)$. Then there exists a* *model* $(\Delta, \alpha, \Xi)$ *for* $\mathcal{T}$ *such that*

$$\overline{\alpha}(S_1) \cap \overline{\alpha}(S_2) \neq \emptyset.$$

***Proof.*** We use the construction described in Lemma 6, with the exception of step 3 of the construction of the model $\mathcal{M}_S$ for the subtree $\mathcal{T}_S$ with root $S$, where we do the following:

- We choose an element $x_1$ in $\overline{\alpha}_1(S_1)$ in the model $\mathcal{M}_{S'_1}$ built for $\mathcal{T}_{S'_1}$ (we know there exists one such $x_1$ because $\mathsf{CInf}(S_1) > 0$)
- Analogously, we choose an element $x_2$ in $\overline{\alpha}_2(S_2)$ in the model $\mathcal{M}_{S'_2}$ build for $\mathcal{T}_{S'_2}$
- We choose the bijections $\sigma_{Q_1}$ and $\sigma_{Q_2}$ in step 3 such that $\sigma_{Q_1}(x_1) = \sigma_{Q_2}(x_2)$. ∎

LEMMA 19 (Non-empty intersection (2)). *Let $S_1$ and $S_2$ be nodes in a weakly consistent i-tree $\mathcal{T}$ is such that*

$$\mathsf{CInf}(S_1) > 0 \wedge S_1 \stackrel{*}{\leadsto} S'_1 \wedge S'_1 \in Q_1 \wedge$$
$$\mathsf{CInf}(S_2) > 0 \wedge S_2 \stackrel{*}{\leadsto} S'_2 \wedge S'_2 \in Q_2$$

*for some $S, S'_1, S'_2 \in \mathbb{S}$, $Q_1, Q_2 \in \mathsf{Split}(S)$ where $Q_1 \neq Q_2$, and $Q_1 \subseteq C$, $Q_2 \subseteq C$ for some $C \in \mathsf{Comp}(S)$ with the property*

$$\mathsf{CSup}(S) < \Sigma(\mathsf{CSup}[C]).$$

*Then there exists a model $(\Delta, \alpha, \Xi)$ for $\mathcal{T}$ such that*

$$\overline{\alpha}(S_1) \cap \overline{\alpha}(S_2) \neq \emptyset.$$

***Proof.*** We use the construction described in the proof of Lemma 6, except for the step 1 of the construction of the model $\mathcal{M}_S$ for the subtree $\mathcal{T}_S$ with root $S$, for which we do the following. From $K(S) \leq \mathsf{CSup}(S) < \Sigma(\mathsf{CSup}[Q])$, we conclude that the choice of cardinalities $t_Q$ for $Q \in \mathbb{C}$ in the proof of Lemma 6 is such that $\Sigma_{Q \in \mathbb{C}} t_Q > K(S)$, by considering two cases.

1. There exists $Q \in \mathbb{C}$ such that

$$t_Q = \min(K(S), \Sigma(\mathsf{CSup}[Q])) = K(S).$$

Then we choose $Q_i \in \{Q_1, Q_2\}$ such that $Q_i \neq Q$. Since $\mathsf{CInf}(S_1) > 0 \wedge S_1 \stackrel{*}{\leadsto} S'_1 \leadsto S$, repeatedly applying $\mathcal{C}_4$ and using $\mathcal{C}_5$, we have

- $\Sigma(\mathsf{CSup}[Q_i]) \geq \mathsf{CSup}(S'_i) \geq \mathsf{CInf}(S'_i) \geq \mathsf{CInf}(S_i) > 0$
- $K(S) \geq \mathsf{CInf}(S) \geq \mathsf{CInf}(S_i) > 0$

As a consequence $t_{Q_i} = \min(K(S), \Sigma(\mathsf{CSup}[Q_i])) > 0$ and $\Sigma_{Q \in \mathbb{C}} t_Q \geq t_Q + t_{Q_i} > K(S)$.

2. For all $Q \in \mathbb{C}$, $t_Q = \Sigma(\mathsf{CSup}[Q])$). Then $\Sigma_{Q \in \mathbb{C}} t_Q = \Sigma(\mathsf{CSup}[C]) > \mathsf{CSup}(S) \geq K(S)$.

Because $\Sigma_{Q \in \mathbb{C}} t_Q > K(S)$, we can apply Remark 2 and choose one element $x_1$ in $\overline{\alpha}_1(S_1)$ in the model $\mathcal{M}_{S'_1}$ built for $\mathcal{T}_{S'_1}$, and an element $x_2$ in $\overline{\alpha}_2(S_2)$ in the model built for $\mathcal{T}_{S'_2}$, and decide to merge these elements in the step 1 of the construction. We know that such elements $x_1, x_2$ exist because $\mathsf{CInf}(S_1) > 0$ and $\mathsf{CInf}(S_2) > 0$. ∎

LEMMA 20 (Isolated element). *Let $\mathcal{T}$ be a weakly consistent i-tree and $(Q_1, \leadsto)$ a subtree of $(\mathbb{S}, \leadsto)$ with the same root $S_R$ as $\mathcal{T}$, such that for all $S \in Q_1$ all the following conditions hold:*

1. $\mathsf{CInf}(S) > 0$
2. $\forall C \in \mathsf{Comp}(S).\ \exists^{=1} Q \in \mathsf{Split}(S).\ Q \subseteq C \wedge |Q \cap Q_1| = 1$
3. $\forall Q \in \mathsf{Split}(S).\ |Q \cap Q_1| \neq 1 \Rightarrow$
$(|Q \cap Q_1| = 0 \wedge \Sigma(\mathsf{CInf}[Q]) < \mathsf{CInf}(S)).$

*Then we can construct a model for $\mathcal{T}$ such that*

$$\exists x \in \overline{\alpha}(R).\ \forall S \in \mathbb{S}.\ (x \in \overline{\alpha}(S) \Leftrightarrow S \in Q_1)$$

***Proof.*** We use a variation of the construction in the proof of Lemma 6. We apply the assumptions about the subtree $Q_1$ to show that we always have enough "slack" to avoid merging one specific element from $Q_1$ with the elements of neighbors. We being by describing a slightly modified Step 1 of the proof of Lemma 6.

**Step 1' (for nodes of $Q_1$).** Consider the root $S_R \in Q_1$. Let $C \in \mathsf{Comp}(S_R)$, and $\mathbb{C} \subseteq \mathsf{Split}(S_R)$ such that $C = \cup \mathbb{C}$. By construction, there exists a unique son $S' \in Q_1$ of $S_R$ and a corresponding split view $Q'$ such that $S' \in Q'$ and $\mathbb{C} = \{Q'\} \uplus \mathbb{C}_0$. We define $t_{Q'} = \min(k, \Sigma(\mathsf{CSup}[Q']))$ and for each $Q_0 \in \mathbb{C}_0$, we define $t_{Q_0} \stackrel{def}{=} \min(k-1, \Sigma(\mathsf{CSup}[Q_0]))$. For each $Q_0 \in \mathbb{C}_0$, since $k \geq \mathsf{CInf}(S_R)$ by choice of $k$ and $\mathsf{CInf}(S_R) > \Sigma(\mathsf{CInf}[Q_0])$ by hypothesis 3 on $\mathcal{T}$, we have $\Sigma(\mathsf{CInf}[Q_0]) \leq k-1$, so

$$\Sigma(\mathsf{CInf}[Q_0]) \leq t_{Q_0} \leq \Sigma(\mathsf{CSup}[Q_0]) \qquad (H1)$$

and the property (H1) also holds for $t_{Q'}$, because $t_{Q'}$ is defined as in Lemma 6.

By definition of $t$ we clearly have $\max_{Q \in \mathbb{C}} t_Q \leq k$. We next show $\sum_{Q \in \mathbb{C}} t_Q \geq k$ by considering the following cases.

- $t_{Q'} = k$. Then the claim is obvious.
- For all $Q \in \mathbb{C}$ we have $t_Q = \Sigma(\mathsf{CSup}[Q])$. The claim follows from $\mathcal{C}_3$ and the choice of $k$ because $\Sigma t_Q \geq \Sigma(\mathsf{CSup}[C]) \geq \mathsf{CSup}(S_R) \geq k$.
- There exists $Q_0 \in \mathbb{C}_0$ such that $t_{Q_0} = k - 1$. Using $\Sigma(\mathsf{CSup}[Q']) \geq \mathsf{CSup}(S') > 0$ and $k \geq \mathsf{CInf}(S_R) > 0$, we obtain $t_{Q'} > 0$, so

$$\sum_{Q \in \mathbb{C}} t_Q \geq t_{Q_0} + t_{Q'} \geq (k-1) + 1 \geq k.$$

We finally obtain

$$\max_{Q \in \mathbb{C}} t_Q \leq k \leq \sum_{Q \in \mathbb{C}} t_Q \qquad (H2)$$

By definition of all $t_{Q_0}$ we then have

$$\forall Q_0 \in \mathbb{C}_0.\ t_{Q_0} < k \qquad (H3)$$

According to Remark 2, H3 allows us to choose an element of the model $\mathcal{M}_{S'}$ constructed for the subtree $\mathcal{T}_{S'}$ and decide not to merge it with any other element. This observation allows us to recursively enforce $x \in \overline{\alpha}(S) \iff S \in Q_1$.

Indeed, consider a node $S_R \in Q_1$ and let $\{S_R^1, \ldots, S_R^p\} = \mathsf{Sons}(S_R) \cap Q_1$ be its sons in $Q_1$. For each $i$, we can then recursively ensure $x_i \in \overline{\alpha}(S) \iff S \in Q_1$ for each $S$ in the $S_R^i$ subtree i.e. for each $S$ for which $S \stackrel{*}{\leadsto} S_R^i$. By definition of $Q_1$, each $S_R^i$ is in a different complete view, so we can apply bijection to the submodels (Remark 4) and let $\sigma_1(x_1) = \ldots = \sigma_p(x_p) = x$. We ensure that $x$ does not belong to any subtree rooted at a node $S_0 \in \mathsf{Sons}(S_R) \setminus Q_1$, using Remark 2 to make sure that $x$ is not merged with any of the elements of $\overline{\alpha}(S_0)$, which is possible thanks to H3. Finally, for the base case, when $S$ has no sons, we pick $x$ to be a fresh element, which is possible by assumption 3 on the subtree $Q_1$, as noted in Remark 3. ∎

### A.4.2 Links between weak and strong consistency

**Lemma 8 (Bounds Refinement)** *Let $\mathcal{T}$ be a strongly consistent i-tree, $S \in \mathbb{S}$, $i, s$ such that $\mathsf{CInf}(S) \leq i \leq s \leq \mathsf{CSup}(S)$, let $\mathcal{T}' = \mathcal{T}[\mathsf{CInf}(S) \leftarrow i, \mathsf{CSup}(S) \leftarrow s]$ and $\mathcal{T}'_{\mathsf{NF}} = \mathbf{R}^w_{\mathsf{NF}}(\mathcal{T}')$. Then 1) $\mathcal{T}'_{\mathsf{NF}} \neq \perp_d$, 2) $\mathcal{T}'_{\mathsf{NF}} \models \mathcal{T}$, and 3) if $\neg(S \stackrel{*}{\leadsto} S_0)$, then*

$$(\mathsf{CInf}(S_0), \mathsf{CSup}(S_0)) = (\mathsf{CInf}'_{\mathsf{NF}}(S_0), \mathsf{CSup}'_{\mathsf{NF}}(S_0)).$$

***Proof.***

1. We prove this result by induction on the depth of $S$ in the tree $(\mathbb{S}, \leadsto)$. The key step of this proof is to show that the application of $\mathsf{UpSup}$ and/or $\mathsf{UpInf}$ to the father $S'$ of $S$ do not produce a situation where $a_5$ holds in the resulting diagram $\mathcal{T}''$ (and therefore the rule Error is not applicable in $\mathcal{T}''$). We distinguish three different cases:

- When UpSup and UpInf are both applicable to this node $S'$ we have $\mathsf{CInf}''(S') \leq \mathsf{CSup}''(S')$ because for $C \in \mathsf{Comp}(S'), Q \in \mathsf{Split}(S')$ such that $Q \subseteq C, Q = \{S\} \uplus Q_0, C = \{S\} \uplus C_0$ and $\mathcal{T}' \xrightarrow[\mathbf{UpSup}]{S',C} \xrightarrow[\mathbf{UpInf}]{S',Q} \mathcal{T}''$ we have

$$\begin{aligned}
\mathsf{CInf}''(S') &= \Sigma(\mathsf{CInf}'[Q_0]) + \mathsf{CInf}'(S) & \text{(by } b_4) \\
&= \Sigma(\mathsf{CInf}'[Q_0]) + \mathsf{CSup}'(S) & (i \leq s) \\
&\leq \Sigma(\mathsf{CSup}'[Q_0]) + \mathsf{CSup}'(S) & \text{(by } \mathcal{C}_5) \\
&\leq \Sigma(\mathsf{CSup}'[C_0]) + \mathsf{CSup}'(S) & (Q_0 \subseteq C_0) \\
&= \mathsf{CSup}''(S') & \text{(by } b_3)
\end{aligned}$$

- When only UpSup is applicable to this node $S'$ we have $\mathsf{CInf}''(S') \leq \mathsf{CSup}''(S')$ because for $C \in \mathsf{Comp}(S'), C = \{S\} \uplus C_0$ and $\mathcal{T}' \xrightarrow[\mathbf{UpSup}]{S',C} \mathcal{T}''$ we have

$$\begin{aligned}
\mathsf{CSup}''(S') &= \Sigma(\mathsf{CSup}'[C_0]) + \mathsf{CSup}'(S) & \text{(by } b_3) \\
&\geq \Sigma(\mathsf{CSup}'[Q_0]) + \mathsf{CInf}'(S) & (i \leq s) \\
&\geq \mathsf{CInf}'(S') & \text{(by } \mathcal{C}_6) \\
&= \mathsf{CInf}''(S')
\end{aligned}$$

- When only UpInf is applicable to this node $S'$ we have $\mathsf{CInf}''(S') \leq \mathsf{CSup}''(S')$ because for $Q \in \mathsf{Split}(S'), Q = \{S\} \uplus Q_0$ and $\mathcal{T}' \xrightarrow[\mathbf{UpInf}]{S',Q} \mathcal{T}''$ we have

$$\begin{aligned}
\mathsf{CInf}''(S') &= \Sigma(\mathsf{CInf}'[Q_0]) + \mathsf{CInf}'(S) & \text{(by } b_4) \\
&\leq \Sigma(\mathsf{CInf}'[Q_0]) + \mathsf{CSup}'(S) & (i \leq s) \\
&\leq \mathsf{CSup}'(S') & \text{(by } \mathcal{C}_7) \\
&= \mathsf{CSup}''(S')
\end{aligned}$$

2. Follows easily from the hypothesis $\mathsf{CInf}(S) \leq i \leq s \leq \mathsf{CSup}(S)$ and the fact that $\mathbf{R}_{\mathsf{NF}}^{\mathsf{w}}$ is semantics preserving.

3. It is enough to notice that only rules UpInf and UpSup are used when applying $\mathbf{R}_{\mathsf{NF}}^{\mathsf{w}}$, and these rules are applied in the bottom-up direction. ∎

**Lemma 9 (Parallel Bounds Refinement (1))** *Let $\mathcal{T}$ be a strongly consistent i-tree, and $(Q_0, \rightsquigarrow)$ a subtree of $\mathcal{T}$ which has the same root as $\mathcal{T}$ and is such that*

- *The nodes of $Q_0$ are pairwise independent, that is*

$$\forall S_1, S_2 \in Q_0. \neg(\mathsf{disj}_\mathcal{T}^*(S_1, S_2))$$

*Then the i-tree $\mathcal{T}'$ defined by*

$$\mathcal{T}' \stackrel{def}{=} \mathcal{T}[\forall S \in Q_0 : \mathsf{CInf}(S) \leftarrow \mathsf{CSup}(S)]$$

*is such that his $\mathcal{R}^w$ normal form $\mathcal{T}'_{\mathsf{NF}} \stackrel{def}{=} \mathbf{R}_{\mathsf{NF}}^{\mathsf{w}}(\mathcal{T}')$ satisfies*

1. $\mathcal{T}'_{\mathsf{NF}} \neq \bot_d$
2. $\mathcal{T}'_{\mathsf{NF}} \models \mathcal{T}$
3. $\forall S \in Q_0. \forall Q \in \mathsf{Split}'_{\mathsf{NF}}(S). Q \cap Q_0 = \emptyset \Rightarrow \Sigma(\mathsf{CInf}'_{\mathsf{NF}}[Q]) < \mathsf{CInf}'_{\mathsf{NF}}(S)$

**Proof.** If we apply $[\mathsf{CInf}(S') \leftarrow \mathsf{CSup}(S')]$ to every node $S'$ of $Q_0$ starting from the root to the leaves, we always maintain $\mathcal{C}_1, \mathcal{C}_2, \mathcal{C}_3$ because $\Phi$ and $\mathsf{CSup}$ are never modified. We also maintain $\mathcal{C}_4$ because for each $S \in Q_0$ and each view $Q \in \mathsf{Split}(S)$ such that there exists $S' \in Q \cap Q_0$, by $\neg\mathsf{disj}_\mathcal{T}^*(S_1, S_2)$, we know that $S'$ is the only modified node, and

$$\begin{aligned}
\mathsf{CInf}'(S) &= \mathsf{CSup}(S) \\
&\geq \Sigma(\mathsf{CInf}[Q - \{S'\}]) + \mathsf{CSup}(S') & \text{(by } \mathcal{C}_7) \\
&= \Sigma(\mathsf{CInf}'[Q])
\end{aligned}$$

Therefore, $\mathcal{T}'$ is already in normal form, so $\mathcal{T}'_{\mathsf{NF}}$ is identical to $\mathcal{T}'$ and is clearly distinct from $\bot_d$, proving condition 1. Condition 2 holds because $\mathcal{T}' \models \mathcal{T}$ because the cardinality bounds in $\mathcal{T}'$ are at least as strong as in $\mathcal{T}$. Condition 3 holds because

$$\begin{aligned}
\Sigma(\mathsf{CInf}'[Q]) &= \Sigma(\mathsf{CInf}[Q]) & \text{(because } Q \cap Q_0 = \emptyset) \\
&< \mathsf{CSup}(S) & \text{(by } \mathcal{C}_8) \\
&= \mathsf{CInf}'(S) & \text{(by definition of } \mathcal{T}')
\end{aligned}$$
∎

LEMMA 21 (Parallel Bounds Refinement (2)). *Let $\mathcal{T}$ be a strongly consistent i-tree and $S_1, S_2, S_1', S_2', S \in \mathbb{S}$ and $C, Q$ such that*

$$C \in \mathsf{Comp}(S), Q_1, Q_2 \in \mathsf{Split}(S)$$
$$S_1 \stackrel{*}{\rightsquigarrow} S_1' \wedge S_1' \in Q_1 \wedge Q_1 \subseteq C$$
$$S_2 \stackrel{*}{\rightsquigarrow} S_2' \wedge S_2' \in Q_2 \wedge Q_2 \subseteq C$$
$$Q_1 \neq Q_2$$

*Define*

$$\begin{aligned}
\mathcal{T}' &\stackrel{def}{=} \mathcal{T}[\forall S', S_1 \stackrel{*}{\rightsquigarrow} S' \stackrel{*}{\rightsquigarrow} S_1' : \mathsf{CInf}(S') \leftarrow \mathsf{CSup}(S')] \\
&\phantom{\stackrel{def}{=}} [\forall S', S_2 \stackrel{*}{\rightsquigarrow} S' \stackrel{*}{\rightsquigarrow} S_2' : \mathsf{CInf}(S') \leftarrow \mathsf{CSup}(S')] \\
\mathcal{T}'_{\mathsf{NF}} &\stackrel{def}{=} \mathbf{R}_{\mathsf{NF}}^{\mathsf{w}}(\mathcal{T}') \\
\mathcal{T}'' &\stackrel{def}{=} \mathbf{R}_{\mathsf{NF}}^{\mathsf{w}}(\mathcal{T}'_{\mathsf{NF}}[\mathsf{CSup}''(S) \leftarrow \mathsf{CInf}'_{\mathsf{NF}}(S)])
\end{aligned}$$

*Then $\mathcal{T}'' \neq \bot_d$, $\mathcal{T}'' \models \mathcal{T}$, and $\mathsf{CSup}''(S) < \Sigma(\mathsf{CSup}''[C])$.*

**Proof.** As in the proof of Lemma 9, weak consistency conditions hold in $\mathcal{T}'$ for all nodes in $S'$ such that $S_1 \stackrel{*}{\rightsquigarrow} S_1'$ or $S_2 \stackrel{*}{\rightsquigarrow} S_2'$. Therefore, the only rewrite step that mat be applicable in $\mathcal{T}'$ is the application of **UpInf** to $S$. This application may lead to applications of other instances of **UpInf**, but the proof of Lemma 8 shows that this process will result in a weakly consistent i-tree, so $\mathcal{T}'_{\mathsf{NF}} \neq \bot_d$. Moreover, the process of computing $\mathbf{R}_{\mathsf{NF}}^{\mathsf{w}}(\mathcal{T}'_{\mathsf{NF}}[\mathsf{CSup}''(S) \leftarrow \mathsf{CInf}'_{\mathsf{NF}}(S)])$ is identical to applying Lemma 8 to $\mathcal{T}'$ with bounds $i = s = \mathsf{CInf}'_{\mathsf{NF}}(S)$, and therefore leads to a weakly consistent i-tree $\mathcal{T}''$, so $\mathcal{T}'' \neq \bot_d$.

The condition $\mathcal{T}'' \models \mathcal{T}$ follows because $\mathbf{R}_{\mathsf{NF}}^{\mathsf{w}}$ is semantics-preserving, and the updates of trees only shrink the bounds on nodes, so they convert a diagram into a stronger one.

To prove $\mathsf{CSup}''(S) < \Sigma(\mathsf{CSup}''[C])$, observe first that $\mathsf{CSup}''(S) = \mathsf{CInf}'_{\mathsf{NF}}(S)$ by definition of $\mathcal{T}''$, and $\Sigma(\mathsf{CSup}''[C]) = \Sigma(\mathsf{CSup}[C])$ because $\mathsf{CSup}$ does not change for any ancestors of $S$. Therefore, it suffices to show

$$\mathsf{CInf}'_{\mathsf{NF}}(S) < \Sigma(\mathsf{CSup}[C])$$

We prove this condition by distinguishing two cases.

1. **UpInf** is not applicable to $S$. Then $\mathsf{CInf}'_{\mathsf{NF}}(S) = \mathsf{CInf}(S)$ and the condition follows by $\mathcal{C}_9$.
2. **UpInf** is applicable to $S$. Then for some $a, b$ where $\{a, b\} = \{1, 2\}$ we have

$$\begin{aligned}
\mathsf{CInf}'_{\mathsf{NF}}(S) &= \Sigma(\mathsf{CInf}'[Q_a]) \\
&< \Sigma(\mathsf{CInf}'[Q_a]) + \Sigma(\mathsf{CInf}'[Q_b]) \\
&\leq \Sigma(\mathsf{CInf}'[C]) \\
&\leq \Sigma(\mathsf{CSup}[C])
\end{aligned}$$
∎

### A.4.3 Completeness of the algorithm Subsumes

**Theorem 4** *Let $\mathcal{T}$ be a strongly consistent i-tree and let $\mathsf{H}_A^\mathcal{T}$ for atomic formula $A$ be as defined in Figure 12. Then $\mathsf{H}_A^\mathcal{T}$ if and only if $\mathcal{T} \models A$.*

The ($\Rightarrow$) direction of Theorem 4 is trivial by the semantics of i-diagrams. For ($\Leftarrow$) direction we prove the following characteriza-

tions:

$$S \neq \emptyset_d \Rightarrow \exists \mathcal{M}.\ \overline{\alpha}(S) \neq \emptyset$$
$$k \in [\mathsf{CInf}(S), \mathsf{CSup}(S)] \Rightarrow \exists \mathcal{M}.\ |\overline{\alpha}(S)| = k$$
$$\neg(\mathbf{Included}(S_0, C, \mathcal{T})) \Rightarrow \exists \mathcal{M}.\ \overline{\alpha}(S_0) \not\subseteq \bigcup \overline{\alpha}[C]$$
$$S_1 \neq \emptyset_d \wedge \neg(S_1 \overset{*}{\leadsto} S_2) \Rightarrow \exists \mathcal{M}.\ \overline{\alpha}(S_1) \not\subseteq \overline{\alpha}(S_2)$$
$$S_1 \neq S_2 \Rightarrow \exists \mathcal{M}.\ \overline{\alpha}(S_1) \neq \overline{\alpha}(S_2)$$
$$\emptyset_d \notin \{S_1, S_2\} \wedge \neg\mathsf{disj}^*_\mathcal{T}(S_1, S_2) \Rightarrow \exists \mathcal{M}.\ \overline{\alpha}(S_1) \cap \overline{\alpha}(S_2) \neq \emptyset$$
$$+P \notin \Phi(S) \Rightarrow \exists \mathcal{M}.\ \overline{\alpha}(S) \not\subseteq \Xi(P)$$
$$-P \notin \Phi(S) \Rightarrow \exists \mathcal{M}.\ \overline{\alpha}(S) \not\subseteq \Xi(P)^c$$

where $\mathcal{M}$ denotes a model $\mathcal{M} = (\Delta, \alpha, \Xi)$ of $\mathcal{T}$. We next present the remaining lemmas that prove these characterizations (see also Section 7).

**Lemma 12** *If an i-tree $\mathcal{T}$ is strongly consistent, $S_0, \in \mathbb{S}, C \in \mathcal{P}(\mathbb{S})$, and $\mathbf{Included}(S_0, C, \mathcal{T})$ returns false, then*

$$\exists \mathcal{M}.\ \overline{\alpha}(S_0) \not\subseteq \bigcup \overline{\alpha}[C]$$

*Proof.* Let $Q_0$ be the smallest set of nodes such that:

$$\begin{array}{l} S_0 \overset{*}{\leadsto} S \Rightarrow S \in Q_0 \\ S \in Q_0 \wedge Q_1 \in \mathsf{Comp}(S) \wedge \\ S_1 \in Q_1 \wedge \neg\mathbf{Incl}(S_1, C) \end{array} \Rightarrow S_1 \in Q_0$$

By definition of **Incl** we know that $Q_0 \cap C = \emptyset$.

$Q_0$ is tree-shaped by construction, but may contain two nodes which are explicitly disjoint. We compute a subtree $Q_1$ of $Q_0$, by starting from the root and keeping each time at most one son for each complete view. We also impose that $Q_1$ contains $S$, by avoiding to cut the branch which leads to $S$.

We then define

$$\mathcal{T}' = \mathbf{R}^w_{\mathsf{NF}}(\mathcal{T}[\forall S \in Q_1 : \mathsf{CInf}(S) \leftarrow \mathsf{CSup}(S)])$$

According to Lemma 9 (Parallel Bounds Refinement) we have $\mathcal{T}' \neq \perp_d$.

We then apply Lemma 20 to construct a model for the weakly consistent i-tree $\mathcal{T}'$ such that

$$\forall S' \in \mathbb{S}.\ (x \in \overline{\alpha}(S') \Leftrightarrow S' \in Q_1)$$

for some element $x \in \overline{\alpha}(S)$. Because $S \in Q_1$, we have $x \in \overline{\alpha}(S)$. Because $C \cap Q_1 = \emptyset$, we have $x \notin \bigcap \overline{\alpha}[C]$. ∎

**Lemma 15** *If an i-tree $\mathcal{T}$ is strongly consistent, then for all $S \in \mathbb{S}$ and $P \in \mathsf{PN}$ we have*

$$(+P) \notin \Phi(S) \Rightarrow \exists \mathcal{M}.\ \mathcal{M} \models \mathcal{T} \wedge \overline{\alpha}_\mathcal{M}(S) \not\subseteq \Xi(P)$$
$$(-P) \notin \Phi(S) \Rightarrow \exists \mathcal{M}.\ \mathcal{M} \models \mathcal{T} \wedge \overline{\alpha}_\mathcal{M}(S) \not\subseteq \Xi(P)^c$$

*Proof.* Let $S \in \mathbb{S}$ be such that $(+P) \notin \Phi(S)$. We define $Q_{+P} \overset{def}{=} \{S' \in \mathbb{S} | (+P) \in \Phi(S')\}$. Using $\mathcal{C}_{10}$ we show that $\mathbf{Included}(S, Q_{+P}, \mathcal{T})$ cannot return true. Then, there exists a model such that $\overline{\alpha}(S) \not\subseteq (\bigcup \overline{\alpha}[Q_{+P}])$. We then change the model by redefining $\Xi'$ by:

$$\forall P \in \mathsf{PN}.\ \Xi'(P) = \bigcup \{\overline{\alpha}(S') | S' \in Q_{+P}\}$$

to ensure that $\Xi'(P) \not\subseteq \overline{\alpha}(S)$. The case of $(-P) \notin \Phi(S)$ is analogous by taking a model such that $\overline{\alpha}(S) \not\subseteq (\bigcup \overline{\alpha}[Q_{-P}])$ and redefining $\Xi'$ by:

$$\forall P \in \mathsf{PN}.\ \Xi'(P) = \Delta - \bigcup \{\overline{\alpha}(S') | S' \in Q_{-P}\}$$

∎

**Lemma 16** *If an i-tree $\mathcal{T}$ is strongly consistent, then for all $S_1, S_2 \in \mathbb{S}$ such that $S_1 \neq \emptyset, S_2 \neq \emptyset$ we have*

$$\neg(\mathsf{disj}^*_\mathcal{T}(S_1, S_2)) \Rightarrow \exists \mathcal{M}.\ \overline{\alpha}(S_1) \cap \overline{\alpha}(S_2) \neq \emptyset.$$

*Proof.* Let $S_1, S_2 \in \mathbb{S} \setminus \{\emptyset_d\}$ such that $\neg(\mathsf{disj}^*_\mathcal{D}(S_1, S_2))$. If $S_1 = S_2$ we can find a model $\mathcal{M}$ where $\overline{\alpha}(S_1) = \overline{\alpha}(S_2) \neq \emptyset$ by Lemma 10, in this model $\overline{\alpha}(S_1) \cap \overline{\alpha}(S_2) = \overline{\alpha}(S_2) \neq \emptyset$. Suppose $S_1 \neq S_2$. Define $S'_1, S'_2$, $S_0$ as the unique nodes such that $S_0$ is the least common ancestor of $S_1$ and $S_2$ in $\mathcal{T}$, and $S'_1, S'_2 \in \mathsf{Sons}(S_0)$ are the ancestors of $S_1$ and $S_2$, respectively. We distinguish two cases:

- $S'_1$ and $S'_2$ do not belong to a same complete view of $S_0$. Then apply Lemma 9 to the subtree

$$Q_0 \overset{def}{=} \{S \in \mathbb{S} \mid S_1 \overset{*}{\leadsto} S \vee S_2 \overset{*}{\leadsto} S\}$$

whose nodes are pairwise independent by the hypothesis $\mathsf{disj}^*_\mathcal{T}(S_1, S_2)$. The resulting tree $\mathcal{T}'_{\mathsf{NF}}$ satisfies the hypothesis of Lemma 18, so there exists a model $\mathcal{M} = (\Delta, \alpha, \Xi)$ for $\mathcal{T}'_{\mathsf{NF}}$ such that

$$\overline{\alpha}(S_1) \cap \overline{\alpha}(S_2) \neq \emptyset.$$

$\mathcal{M}$ is also a model of $\mathcal{T}$ because $\mathcal{T}'_{\mathsf{NF}} \models \mathcal{T}$.

- $S'_1$ and $S'_2$ belong to a same complete view $C$. Define

$$\mathcal{T}' \overset{def}{=} \mathbf{R}^w_{\mathsf{NF}}(\mathcal{T}[\forall S', S_1 \overset{*}{\leadsto} S' \overset{*}{\leadsto} S'_1 : \mathsf{CInf}(S') \leftarrow \mathsf{CSup}(S')]$$
$$[\forall S', S_2 \overset{*}{\leadsto} S' \overset{*}{\leadsto} S'_2 : \mathsf{CInf}(S') \leftarrow \mathsf{CSup}(S')])$$
$$\mathcal{T}'' \overset{def}{=} \mathbf{R}^w_{\mathsf{NF}}(\mathcal{T}'[\mathsf{CSup}''(S) \leftarrow \mathsf{CInf}'(S)])$$

By Lemma 21, then $\mathcal{T}'' \models \mathcal{T}$ and $\mathsf{CSup}''(S) < \Sigma(\mathsf{CSup}''[C])$. This last property allows us to apply Lemma 19 and prove the existence of a model $(\Delta, \alpha, \Xi)$ for $\mathcal{T}$ such that $\overline{\alpha}(S_1) \cap \overline{\alpha}(S_2) \neq \emptyset$. ∎

## B. Example transformation to CBAC constraints

We illustrate the idea of the algorithm in Figure 2 through an example of checking the validity of the formula

$$|A \cup B| = |A| + |B| - |A \cap B|$$

that is, checking the unsatisfiability of the formula

$$|A \cup B| \neq |A| + |B| - |A \cap B|$$

This formula has no integer variables initially, so the first step does not apply. Furthermore, all set algebra expressions appear already within cardinality constraints, so step 2 is unnecessary as well, as is step 3 because there are no divisibility constraints.

In step 4, we introduce a non-negative integer variable for each cardinality term, yielding the system:

$$i_{A \cup B} \neq i_A + i_B - i_{A \cap B}$$
$$|\mathbf{1}| = \mathsf{MAXC}$$
$$|A \cup B| = i_{A \cup B}$$
$$|A| = i_A$$
$$|B| = i_B$$
$$|A \cap B| = i_{A \cap B}$$

Because the formula is already in the form of a conjunction, there is no need to non-deterministically guess the conjunction in step 5, and we continue with the current constraints.

In step 6, we non-deterministically replace $i_{A \cup B} \neq i_A + i_B - i_{A \cap B}$ with $i_{A \cup B} + 1 \leq i_A + i_B - i_{A \cap B}$ or $i_A + i_B - i_{A \cap B} + 1 \leq i_{A \cup B}$, and we illustrate the first case (the other case needs to be checked as well unless the satisfying assignment is found in the

first case):
$$i_{A \cup B} + 1 \leq i_A + i_B - i_{A \cap B}$$
$$|\mathbf{1}| = \mathsf{MAXC}$$
$$|A \cup B| = i_{A \cup B}$$
$$|A| = i_A$$
$$|B| = i_B$$
$$|A \cap B| = i_{A \cap B}$$

In step 7, we introduce a slack variable $i_0$ to eliminate the first inequation, and transform the equation to normal form:
$$i_{A \cup B} - i_A - i_B + i_{A \cap B} + i_0 = -1$$
$$|\mathbf{1}| = \mathsf{MAXC}$$
$$|A \cup B| = i_{A \cup B}$$
$$|A| = i_A$$
$$|B| = i_B$$
$$|A \cap B| = i_{A \cap B}$$

In step 8, we identify $v = (\mathsf{MAXC}, i_{A \cup B}, i_A, i_B, i_{A \cap B}, i_0)$ and have $A = [0, 1, -1, -1, 1, 1]$ and $d = (-1)$. We have $m_0 = 1$, $m_1 = 5, n_0 = 6$, and $S = 2$. Therefore, $m = 6$, $n = 6$, and $a = 1$. From these values we compute $M = 6 \cdot (6 \cdot 1)^{11} = 6^{12} < 2^{32}$. Therefore, if there exists a solution to the system of equations, there exists a solution that can be represented by at most 32 bits in binary.

In step 9, we guess non-deterministically a solution vector bounded by $M$ that satisfies the first equation (that is, $Ak = -1$). One such solution is $k = (101, 100, 77, 40, 10, 6)$ (written in decimal notation). For this guessed solution, we generate the CBAC constraint:
$$|\mathbf{1}| = 101$$
$$|A \cup B| = 100$$
$$|A| = 77$$
$$|B| = 40$$
$$|A \cap B| = 10$$

This example CBAC constraint does not have a solution, and neither do the remaining examples generated by the non-deterministic algorithm. The search tree corresponding to the non-deterministic algorithm returns false in all branches, so the formula is unsatisfiable, and its negation is valid.

We note that, in this case, the dimensions of the system of equations are such that the estimate $M$ can be improved if we consider the system of equations where the variables $\mathsf{MAXC}, i_{A \cup B}, i_A, i_B, i_{A \cap B}$ are all substituted into the the original integer part of the QFBAPA problem. In general, this alternative estimate is given by $M' = n'(m'a')^{2m'+1}$ where

$$m' = m_0$$
$$n' = \max(n_0 - m_1, 2^S)$$
$$a' = \max\left(\max_{1 \leq q \leq m_0} |d_q|, \max_{p=1}^{m_0} \max_{q \in R} |a_{pq}|, \max_{p=1}^{m_0} \max(\sum_{\substack{q \in Q \\ a_{pq} > 0}} a_{pq}, \sum_{\substack{q \in Q \\ a_{pq} < 0}} (-a_{pq}))\right)$$

where $Q = \{p_1, \ldots, p_{m_1}\}$ are indices of variables denoting cardinalities (and that are being substituted), and $R = \{1, \ldots, n_0\} \setminus R$. In our example,
$$m' = 1$$
$$n' = 4$$
$$a' = \max(1, 1, 2)$$

and $M' = 16$, so it suffices to use only 4 bits to represent the constants in the resulting CBAC constraints generated in step 10.